\documentclass[14 pt]{extbook}
\usepackage[utf8]{inputenc}
\usepackage[american]{babel}
\usepackage[T1]{fontenc}
\usepackage{palatino}

\usepackage[pdftex]{graphicx}  
\usepackage[usenames,dvipsnames,svgnames]{xcolor}
\usepackage{dcolumn}
\usepackage{comment}
\usepackage{scalerel}
\usepackage{lipsum}
\usepackage{footnote}
\usepackage{amsmath}
\usepackage{amsfonts}
\usepackage{amssymb}
\usepackage{color}
\usepackage[colorlinks,bookmarks=false,citecolor=NavyBlue,linkcolor=Red,urlcolor=blue]{hyperref}
\usepackage[export]{adjustbox}
\usepackage{amsthm}
\usepackage{simplewick}
\usepackage{bbold}
\usepackage{MnSymbol}
\usepackage{physics}
\usepackage{wasysym}
\usepackage[toc]{appendix}
\usepackage{mathrsfs}  
\usepackage{braket}
\usepackage{graphics}
\usepackage{todonotes}
\usepackage{multirow}
\usepackage{algorithm}
\usepackage{algpseudocode}
\usepackage{dsfont}
\usepackage{mdframed}
\usepackage{listings}
\usepackage{caption}

\lstdefinestyle{pythonstyle}{
    language=Python,
    basicstyle=\ttfamily\footnotesize,
    numbers=left,
    numberstyle=\tiny,
    stepnumber=1,
    numbersep=5pt,
    backgroundcolor=\color{gray!10},
    showspaces=false,
    showstringspaces=false,
    breaklines=true,
    frame=single,
    captionpos=b,
    tabsize=3
}

\usepackage{tikz,ifthen}
\usepackage{bbold}
\usepackage{tikz-network}
\usetikzlibrary{patterns,decorations.pathreplacing,calligraphy}
\usetikzlibrary{shapes,arrows.meta,decorations.pathmorphing}

\newcommand{\Sop}{{\hat{S}}}
\newcommand{\Id}{{\hat{\mathbb{1}}}}
\newcommand{\paulix}{{\hat{\sigma}^x}}
\newcommand{\pauliy}{{\hat{\sigma}^y}}
\newcommand{\pauliz}{{\hat{\sigma}^z}}
\newcommand{\pauliplus}{{\hat{\sigma}^+}}
\newcommand{\pauliminus}{{\hat{\sigma}^-}}
\newcommand{\spin}{\sigma}

\title{Physics Without Frontiers\\ \textbf{Quantumandu}  \\ Lecture Notes}
\author{Guglielmo Lami}
\date{May 2024}

\usepackage{pgfplots}
\pgfplotsset{compat=1.18} 

\lstdefinestyle{pythonstyle}{
    language=Python,
    backgroundcolor=\color[RGB]{214,239,255},  
    basicstyle=\ttfamily\small,     
    keywordstyle=\color{blue},      
    commentstyle=\color{olive},     
    stringstyle=\color{red},        
    numbers=left,                  
    numberstyle=\tiny\color{gray},  
    stepnumber=1,                  
    numbersep=5pt,                 
    showstringspaces=false,        
    breaklines=true,               
    breakatwhitespace=true,        
    morekeywords={self}            
}

\begin{document}

\begin{titlepage}
    \centering
    \vspace{-0.7cm}
    \includegraphics[width=0.35\textwidth]{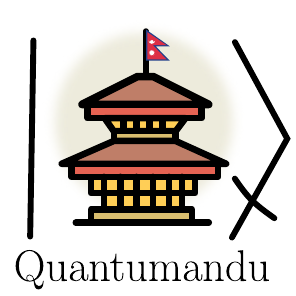}\par\vspace{1cm}
    
    {\Large\bfseries Physics Without Frontiers\par}
    \vspace{0.2cm}
    {\Large\bfseries Quantumandu\par}
    \vspace{0.5cm}
    \begin{center}
    {\Large Beginner's Lecture Notes on \\ Quantum Spin Chains, \\ Exact Diagonalization, \\ Tensor Networks, \\ and Quantum phase transitions \par}
    \end{center}
    \vspace{0.5 cm}
    {\large Guglielmo Lami$^*$, Mario Collura, Nishan Ranabhat \par}
    \vfill
    {\large \today \par}
    \vspace{2 cm}
    *glami@sissa.it
\end{titlepage}

\tableofcontents

\chapter*{Authors' Note}
These notes were prepared on the occasion of the Summer School ``\href{https://indico.ictp.it/event/10721}{Quantumandu}'', which was held at Tribhuvan University (Kathmandu, Nepal) from 25 July to 31 July 2024. The school was part of the Physics Without Frontiers program of the ICTP. These notes are designed with an educational purpose, primarily intended for readers who encounter the physics of strongly correlated many-body systems for the first time. They focus on numerical methods, such as exact diagonalization, and also include a brief introduction to tensor network methods. As these are lecture notes, they may contain errors or typos. Readers are kindly encouraged to report any issues to the authors for correction.

\vspace{3 cm}

\begin{figure*}[h!]
\centering
\includegraphics[width=0.9\linewidth]{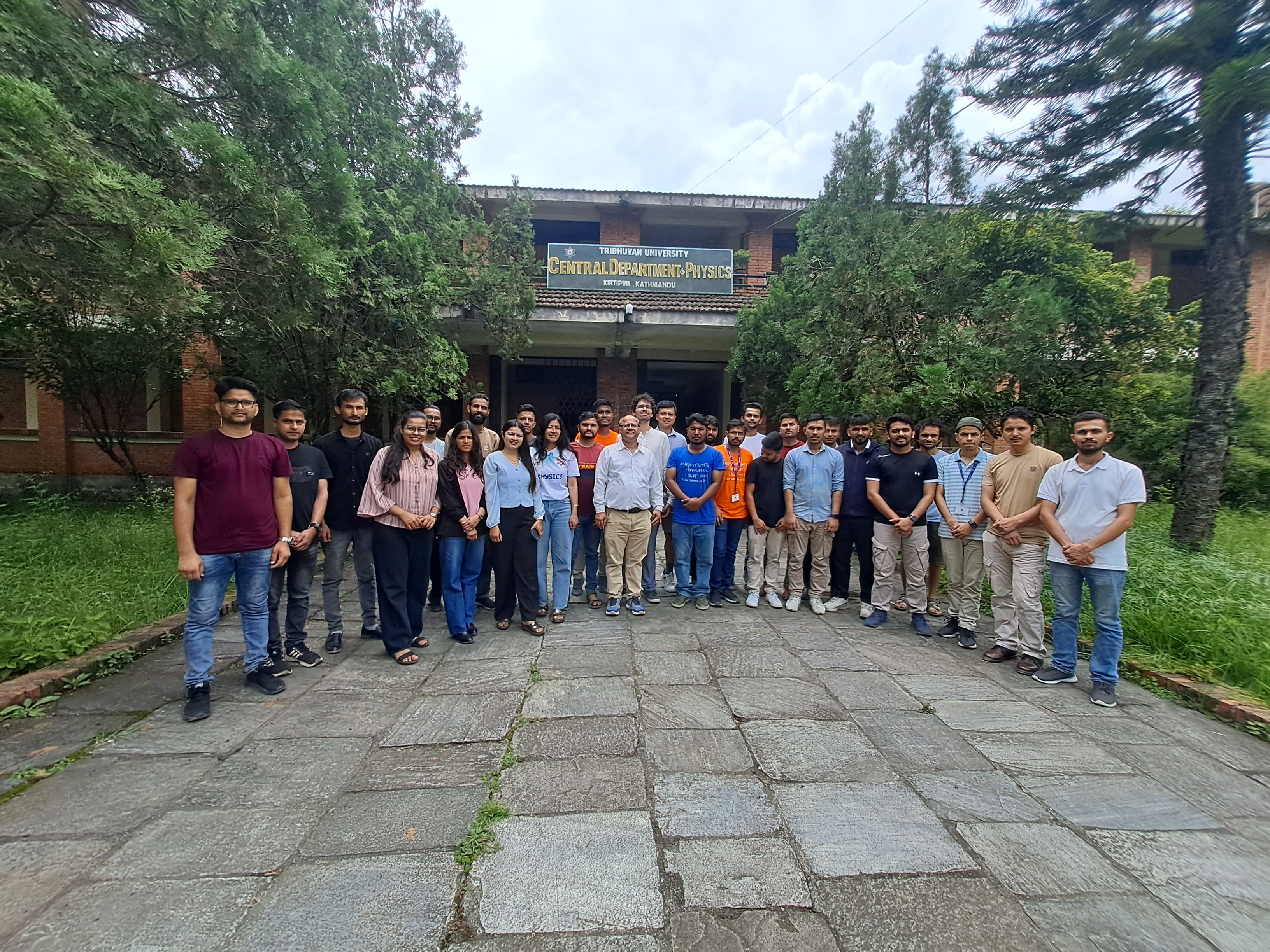}
\end{figure*}
\chapter*{Recommended References}

The notes by Anders W. Sandvik from Boston University, Ref.~\cite{Sandvik_2010}, provide an excellent introduction to quantum spin systems and exact diagonalization methods. They serve as the primary reference for the following lectures. \\ 

\noindent
For an introduction to the main aspects of computational quantum physics, we also recommend the notes by Matthias Troyer, Ref.~\cite{Troyer_2015}. \\

\noindent
For those who feel they have not yet adequately learned quantum mechanics, we suggest the book by Kenichi Konishi and Giampiero Paffuti from University of Pisa, Ref.~\cite{paffuti2009quantum}. It provides a good and rigorous introduction to general quantum mechanics. Also Sakurai's book \cite{sakurai2011modern} is very good. \\

\noindent
For those looking to solidify their understanding of statistical physics and critical phenomena, we recommend the classic book by Kerson Huang, Ref.~\cite{Huang1963StatisticalM2}. Another very good reference are the lecture notes by David Tong from the University of Cambridge, Ref.~\cite{Tong_2012}. \\ 

\noindent
About quantum phase transitions, the most comprehensive reference is the book by Subir Sachdev from Harvard University, Ref.~\cite{Sachdev_2011}. \\

\noindent
The most comprehensive and widely regarded textbook on quantum information and computation is the one by Michael Nielsen and Isaac Chuang Ref.~\cite{Nielsen_chuang_2010}. A very good introduction to these topics is also given by the lecture notes of Giuseppe E. Santoro from SISSA (Trieste), Ref.~\cite{Santoro_2024}. \\    

Regarding Tensor Network methods, an excellent reference is the review by Ulrich Schollwoeck Ref.~\cite{SCHOLLWOCK201196}. We also highlight the recent book Ref.~\cite{Collura2024}.

\chapter{First lecture}
\section{Motivation}
Science explores how things change and evolve through physical transformations. The fundamental concept that modern science uses to model these changes is the \textit{variable}. A variable is a mathematical / philosophical entity that can assume multiple values. The simplest possible variable is a variable $\spin$ that can take only two values. These are known as \textit{binary variables} and are used to model physical systems that can exist in only two distinct \textit{states}. We denote such states as $\ket{\spin}$. An example is a coin that can be observed in states heads or tails.
Other two major examples are the following: 
\begin{itemize}
    \item Computers and other digital devices store and process information in the form of \textit{bits}, which are binary variables. In this context, it is customary to use the notation $\spin \in \{ 0, 1 \}$ for the two possible values.
    \item Elementary particles (as electrons) or atoms can have an inner degree of freedom named \textit{spin}, which is related with their magnetic moment. In the case of spin $1/2$, and when a specific geometrical axis is chosen, the spin can be observed in two states: $\ket{\spin} \in  \{ \ket{\uparrow},\ket{\downarrow} \}$. We will equivalently use the notation $\ket{+1}$, $\ket{-1}$ for the two states respectively (and therefore $\spin$ is $\pm 1$).
\end{itemize}
Both cases are of immense importance in modern science. Bits form the foundation of information and computational science, while spins play a crucial role in statistical physics, condensed matter physics, combinatorics, optimization problems, neural network theory.
In these lectures we will mostly focus on the latter case, as we will consider systems of many interacting spins.

\section{Classical models of spins}
We consider a physical system consisting on $N$ spins $\spin_i$, with index $i$ labelling spins and
taking values $1,2, ... \, N$. The state of the system is specified by the state of each individual spin. We will indicate it in the following equivalent ways
\begin{equation}
    \ket{\pmb{\spin}} \equiv \ket{\spin_1, \spin_2, ... \spin_N} = \ket{\spin_1} \ket{\spin_2} ... \ket{\spin_N} \, ,
\end{equation}
which are abbreviated ways of saying that: spin $1$ is in state $\ket{\spin_1}$, spin $2$ is in state $\ket{\spin_2}$, etc. Notice that there are in total $2^{N}$ possible states $\ket{\pmb{\spin}}$. For instance if $N=3$, the list of all possible states is:
\begin{table}[htbp]
\centering
\begin{minipage}{0.25\linewidth}
\centering
\begin{tabular}{c} 
$\ket{\uparrow \uparrow \uparrow}$ \\ 
\hline
$\ket{\uparrow \uparrow \downarrow}$ \\ 
\hline
$\ket{\uparrow \downarrow \uparrow}$ \\ 
\hline
$\ket{\uparrow \downarrow \downarrow}$ \\ 
\end{tabular}
\end{minipage}%
\begin{minipage}{0.25\linewidth}
\centering
\begin{tabular}{c} 
$\ket{\downarrow \uparrow \uparrow}$ \\ 
\hline
$\ket{\downarrow \uparrow \downarrow}$ \\ 
\hline
$\ket{\downarrow \downarrow \uparrow}$ \\ 
\hline
$\ket{\downarrow \downarrow \downarrow}$ \\
\end{tabular}
\end{minipage}
\end{table}    

To enumerate such states is very convenient to use the convention $0,1$, mapping states $\ket{\pmb{\spin}}$ in strings of $N$ bits. This mapping is very useful, because it provides a one to one mapping from states $\ket{\pmb{\spin}}$ to \textit{binary numbers} from $0$ to $2^{N}-1$, as follows 
\begin{table}[htbp]
\centering
\begin{minipage}{0.35\linewidth}
\centering
\begin{tabular}{c|c|c} 
$\ket{\uparrow \uparrow \uparrow}$ & $000$ & $0$ \\ 
\hline
$\ket{\uparrow \uparrow \downarrow}$ & $001$ & $1$ \\ 
\hline
$\ket{\uparrow \downarrow \uparrow}$ & $010$ & $2$ \\ 
\hline
$\ket{\uparrow \downarrow \downarrow}$ & $011$ & $3$ \\ 
\end{tabular}
\end{minipage}%
\begin{minipage}{0.35\linewidth}
\centering
\begin{tabular}{c|c|c} 
$\ket{\downarrow \uparrow \uparrow}$ & $100$ & $4$ \\ 
\hline
$\ket{\downarrow \uparrow \downarrow}$ & $101$ & $5$ \\ 
\hline
$\ket{\downarrow \downarrow \uparrow}$ & $110$ & $6$ \\ 
\hline
$\ket{\downarrow \downarrow \downarrow}$ & $111$ & $7$ \\
\end{tabular}
\end{minipage}
\end{table}

\subsection{The Ising Hamiltonian}
After defining the system's set of states, also known as the phase space, the next step involves describing the system's physics. This means assigning an \textit{energy} to each state, represented by a function $H(\pmb{\spin})$ that yields a real value of energy for any configuration $\ket{\pmb{\spin}}$. Such a function is commonly referred to as the \textit{Hamiltonian}, and there are numerous physically relevant choices for it, depending on the specific model under study. However, for classical spin systems, the most prevalent Hamiltonian is the following Ising-like Hamiltonian:
\begin{equation}
    H(\pmb{\spin}) = - \frac{1}{2} \sum_{i=1}^{N} \sum_{j=1}^{N}  J_{ij} \spin_i \spin_j \, ,
\end{equation}
where $J_{ij} \in \mathbb{R}$ is a set of real parameters physically interpreted as \textit{couplings} between the spins. To understand the meaning of this Hamiltonian, let us consider first the case of positive coupling, i.e.\ $J_{ij} > 0$. In this case:
\begin{itemize}
    \item the total energy increases by $J_{ij}$ if spin $\spin_i$ and $\spin_j$ are unaligned, i.e.\ if $\spin_i \neq \spin_j$. Indeed in this case $\spin_i \spin_j = -1$ and $-J_{ij} \spin_i \spin_j = J_{ij} > 0$;
    \item the total energy decreases by $J_{ij}$ if spin $\spin_i$ and $\spin_j$ are aligned, i.e.\ if $\spin_i = \spin_j$. Indeed in this case $\spin_i \spin_j = +1$ and $-J_{ij} \spin_i \spin_j = - J_{ij} < 0$.
\end{itemize}
Therefore, $J_{ij}$ represents the \textit{energy cost} one has to pay if spins $\spin_i$, $\spin_j$ are unaligned. If instead $J_{ij} < 0$ this amount would be the \textit{energy gain} of having unaligned spins. Thus, positive couplings $J$ tend to favor the alignment of spins, while negative couplings $J$ tend to favor the anti-alignment of spins. The former behavior is typical of systems known as \textit{ferromagnetic}, while the latter is characteristic of systems known as \textit{antiferromagnetic}. 

In general one can also consider a more general case in which spins are also coupled with an \textit{external magnetic field} $g_i \in \mathbb{R}$, which is represented by introducing an extra term in the Hamiltonian:
\begin{equation}\label{eq:hamiltonian_qubo}
    H(\pmb{\spin}) = - \frac{1}{2} \sum_{i=1}^{N} \sum_{j=1}^{N}  J_{ij} \spin_i \spin_j - \sum_{i=1}^N g_i \spin_i \, ,
\end{equation}
The interpretation as magnetic fields comes from the fact that (if $g_i$ is positive): 
\begin{itemize}
    \item the total energy increases by $g_{i}$ if spin $\spin_i$ is unaligned with $g_i$, i.e.\ if $\spin_i=-1$;
    \item the total energy decreases by $g_{i}$ if spin $\spin_i$ is unaligned with $g_i$, i.e.\ if $\spin_i=+1$. 
\end{itemize}

\subsection{Zero temperature and finite temperature}
Given an Hamiltonian of the kind of Eq.~\ref{eq:hamiltonian_qubo}, a very important question, with a plethora of applications, is finding the optimal configuration of spins $\pmb{\spin}$ such that the energy $H(\pmb{\spin})$ is minimized. This problem is often referred to as the Quadratic Unconstrained Binary Optimization (QUBO) problem~\cite{enwiki:1221833336}. QUBO problems are believed to be NP-hard, meaning that solving it typically requires exponential time in the number of spins $N$. This happens because if the couplings $J$ are random, or almost random,
it is generally impossible to simultaneously satisfy all the alignment / unalignment requirements for all $N(N-1)/2$ pairs of spins present. Consequently, the system has no easily found minimum energy solution, and the only way to find the configuration that minimizes the total energy is to inspect all $2^N$ configurations of the system and compute the energy for all of them.

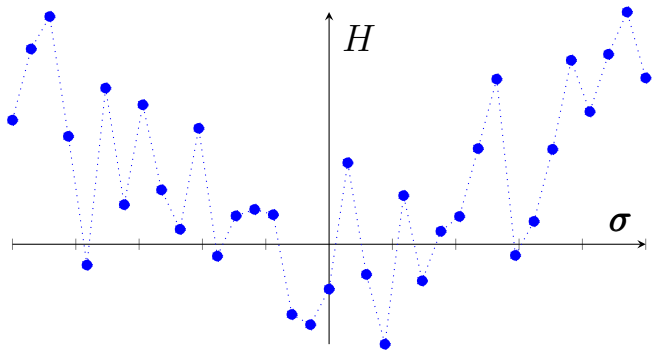
\begin{figure}\label{fig:energy_landscape}
    \centering
    \begin{tikzpicture}
        \begin{axis}[
            xlabel={$\pmb{\sigma}$},
            ylabel={$H$},
            axis lines=middle,
            width=10cm, 
            height=6cm, 
            legend style={at={(1.05,1)}, anchor=north west} 
            xtick=\empty, 
            ytick=\empty,  
            xticklabels=\empty
        ]
            \addplot[
                domain=-10:10, 
                samples=35, 
                mark=*,
                color=blue,
                dotted,
            ]
            {0.27*sin(deg(6*x))-0.29*sin(deg(8*x))+0.015*x^2 + 0.01/(0.1+x^2) + 0.5*rand};  
        \end{axis}
    \end{tikzpicture}
    \caption{The energy landscape of disordered Ising-like Hamiltonians (Eq.\ref{eq:hamiltonian_qubo}) is typically very complex, featuring numerous local minima. Note that the figure serves an illustrative purpose only; in reality, the configuration space $\pmb{\sigma}$ is discrete and multidimensional.}
\end{figure}

However, often physical Hamiltonian have a simpler structure since couplings $J_{ij}$ are just local. This means that spins are considered as to be placed in a certain \textit{lattice}, for instance a square lattice in one or more dimensions, and each spin interacts only with its first nearest neighbors, through 
\begin{equation}
    H(\pmb{\spin}) = - J \sum_{\langle i,j \rangle} \spin_i \spin_j \, ,
\end{equation}
where the symbol $\langle i,j \rangle$ indicates sites of the lattice that are first nearest neighbors.

Another problem of fundamental importance in both physics and optimization algorithms is the ability to determine the probability distribution 
\begin{equation}
    p(\pmb{\spin}) = \frac{1}{Z} e^{-\beta H(\pmb{\spin})}
\end{equation}
where 
\begin{equation}
    Z = \sum_{\pmb{\spin}} e^{-\beta H(\pmb{\spin})} \, .
\end{equation}
$Z$ is a normalization constant, since it guarantees that 
\begin{equation}
    \sum_{\pmb{\spin}} p(\pmb{\spin}) = 1 \, ,
\end{equation}
as expected from a probability distribution. $Z$ is usually named the \textit{partition function}. The distribution $p$ has support over the space of possible configurations $\pmb{\spin}$. This distribution, which is determined by the parameter $\beta = 1/T$, is called the \textit{Boltzmann distribution} (or \textit{Gibbs distribution}) and represents the \textit{thermal fluctuations} of the system when it is brought to a certain temperature $T$. When $T \rightarrow 0$ then $\beta \rightarrow \infty$, meaning that $e^{-\beta H(\pmb{\spin})}$ will increasingly suppress configurations with higher energy, so that for $T=0$, the probability $p$ is actually supported only on the minima of $H$, that is, on the ground states. Conversely, by increasing the temperature $T$ (i.e.\ decreasing $\beta$), the system will progressively be more free to explore configurations with increasingly higher energy. At finite temperature, the physics of a system is therefore determined by the interplay between two factors: the energy value, which determines the Boltzmann factor, and the number of possible configurations corresponding to that energy value. The latter defines the \textit{entropy} of the system. 

A more detailed explanation of these fundamental principles of statistical mechanics will be provided in Lecture \ref{ch:cpt}, within the context of studying phase transitions.

\chapter{Second lecture}
\section{From classical to quantum: the postulates of quantum physics}
Nature is fundamentally non classical. Indeed, quantum mechanics is the correct way of describing the physics of elementary particles~\cite{paffuti2009quantum}. It is therefore natural to ask how to generalize binary variables to the quantum case. 

\subsection{States}
The point is that, while classically one expects the system to always be in one of the two states $\{ \ket{\uparrow},\ket{\downarrow} \}$ even when not observed, in quantum mechanics the system is allowed to be in an arbitrary \textit{linear complex superposition} of the two states when not observed. This means that the system is described by a \textit{wave function} $\ket{\psi}$ which takes the following general form
\begin{equation}
    \ket{\psi} = \psi_0 \ket{\uparrow} + \psi_1 \ket{\downarrow}
\end{equation}
where $\psi_0, \psi_1 \in \mathbb{C}$ are two complex numbers. The mathematical structure of the set of states $\ket{\psi}$ is a complex vector space usually named \textit{Hilbert space} and is denoted as $\mathcal{H}$. $\ket{\uparrow}$ and $\ket{\downarrow}$ are vectors forming a basis of the vector space. An Hilbert space is also equipped with an inner scalar product between states. Given two states $\ket{\psi} = \psi_0 \ket{\uparrow} + \psi_1 \ket{\downarrow}$ and $\ket{\phi}  = \phi_0 \ket{\uparrow} + \phi_1 \ket{\downarrow}$ the inner product is denoted as $\braket{\psi|\phi}$ and is a complex number. It and can be computed as follows
\begin{align}
 \begin{split}
   \braket{\psi|\phi} &= \left( \psi_0^* \bra{\uparrow} + \psi_1^* \bra{\downarrow} \right) \left( \phi_0 \ket{\uparrow} + \phi_1 \ket{\downarrow} \right) = \\ &= \psi_0^* \phi_0 \braket{\uparrow|\uparrow} + \psi_0^* \phi_1 \braket{\uparrow|\downarrow} + \psi_1^* \phi_0 \braket{\downarrow|\uparrow} + \psi_1^* \phi_1 \braket{\downarrow|\downarrow} = \\
   &= \psi_0^* \phi_0 + \psi_1^* \phi_1 \, ,
 \end{split}
\end{align}
where we used the fact that by definition $\braket{\downarrow|\uparrow}=\braket{\uparrow|\downarrow}=0$ (orthogonality of the basis elements) and $\braket{\downarrow|\downarrow}=\braket{\uparrow|\uparrow}=1$ (normalization of the basis elements). 
Notice that 
\begin{equation}
    \braket{\phi|\psi}=(\braket{\psi|\phi})^* \, .
\end{equation}
Given this definition, we can also define what is the \textit{norm} $||\ket{\psi}||$ of a state $\ket{\psi}$. This is defined as:
\begin{equation}
    ||\ket{\psi}|| = \sqrt{\braket{\psi|\psi}} = \sqrt{|\psi_0|^2 + |\psi_1|^2} \, . 
\end{equation}
Vectors $\ket{\uparrow}$ and $\ket{\downarrow}$ forms an \textit{orthonormal basis} of $\mathcal{H}$ because they are orthogonal each other and have norm $1$. We can represent any state as a vector of complex numbers by using its components with respect to this basis (or any other basis). If $ \ket{\psi} = \psi_0 \ket{\uparrow} + \psi_1 \ket{\downarrow}$ we represent it as
\begin{equation}
\left(
\begin{array}{c}
\braket{\uparrow|\psi} \\
\braket{\downarrow|\psi} \\
\end{array}
\right)
= 
\left(
\begin{array}{c}
\psi_0 \\
\psi_1 \\
\end{array}
\right)
\end{equation}
For example the two basis vectors can be represented as 
\begin{equation}
\ket{\uparrow} \longrightarrow
\left(
\begin{array}{c}
1 \\
0 \\
\end{array}
\right) \qquad \ket{\downarrow} \longrightarrow
\left(
\begin{array}{c}
0 \\
1 \\
\end{array}
\right)
\end{equation}
With this representation the inner product $\braket{\psi|\phi}$ can be computed simply by taking the complex conjugate and transpose of the vector $\ket{\psi}$ and computing the standard scalar $\cdot$ product of vectors with $\ket{\phi}$:
\begin{equation}
\braket{\psi|\phi} = 
\left(
\begin{array}{cc}
\psi_0^* & \psi_1^* \\
\end{array}
\right) \cdot \left(
\begin{array}{c}
\phi_0 \\ \phi_1 \\
\end{array}
\right)
\end{equation}

A key assumption of quantum mechanics is that the physical states observed in nature are states having norm 1 and therefore
\begin{equation}
|\psi_0|^2 + |\psi_1|^2 = 1 \, . 
\end{equation}

\subsection{Operators}
Another fundamental postulate asserts that any physical observable $O$ (such as energy, momentum, magnetization, etc.) is associated with a Hermitian operator $\hat{O}$ acting on the Hilbert space $\mathcal{H}$. An operator is a \textit{linear} function $\hat{O}: \mathcal{H} \rightarrow \mathcal{H}$ mapping vectors into vectors. This mapping is denoted by $\hat{O} \ket{\psi} = \hat{O}(\ket{\psi})$.
As for the states, also operators can be represented by using a specific basis. In this case, they takes the form of \textit{matrices} of complex numbers. For instance in the case of a single spin the matrix representing $\hat{O}$ is 
\begin{equation} 
\left(
\begin{array}{cc}
\braket{\uparrow|\hat{O}|\uparrow} & \braket{\uparrow|\hat{O}|\downarrow} \\
\braket{\downarrow|\hat{O}|\uparrow} & \braket{\downarrow|\hat{O}|\downarrow} \\
\end{array}
\right) = \left(
\begin{array}{cc}
O_{00} & O_{01} \\
O_{10} & O_{11} \\
\end{array}
\right)
\end{equation}
where for instance $\braket{\uparrow|\hat{O}|\uparrow}$ means $\braket{\uparrow|\big(\hat{O}|\uparrow}\big)$. The requirement of being Hermitian means that the matrix representing $\hat{O}$ should satisfy $\hat{O}^{\dag} = \hat{O}$, where $\dag$ indicates the Hermitian adjoint, namely conjugate transpose. We have therefore:
\begin{equation} 
\left(
\begin{array}{cc}
O_{00} & O_{01} \\
O_{10} & O_{11} \\
\end{array}
\right) = \left(
\begin{array}{cc}
O_{00}^* & O_{10}^* \\
O_{01}^* & O_{11}^* \\
\end{array}
\right)
\end{equation}
which also implies that diagonal elements $O_{00}=O_{00}^*$, $O_{11}=O_{11}^*$ are real numbers. Another important class of operators (or matrices) are the \textit{unitary operators}. This operator satisfy the property
$\hat{U}^{\dag} = \hat{U}^{-1}$, or equivalently $\hat{U}^{\dag} \hat{U} = \hat{U} \hat{U}^{\dag} = \mathbb{1}$. Another way to define unitary operators is to say that they are operators which preserve the inner product, meaning $\braket{\psi|\psi} = \braket{\psi'|\psi'}$, where $\ket{\psi'} = \hat{U} \ket{\psi}$ and $\ket{\phi'} = \hat{U} \ket{\phi}$.
Unitary operators effectively correspond to a change of basis in the Hilbert space. Indeed, it is clear that $\hat{U}$ maps orthonormal states to orthonormal states. In quantum mechanics, unitary operators are interpreted as possible transformations of a physical system, for instance due to time evolution. 

Hermitian operators, or matrices, $\hat{O}$ can be always diagonalized, meaning that it is possible to find a basis of $\mathcal{H}$ in which the operator appears diagonal. The change of basis is represented by an unitary $\hat{U}$ and the diagonalization is
\begin{equation} 
\hat{U} \hat{O} \hat{U}^{\dag} = \hat{D}
\end{equation}
where $\hat{D} = \text{diag}(\lambda_1, \lambda_2, \lambda_3 ...)$ is a diagonal matrix containing the \textit{eigenvalues} of the matrix $\hat{O}$. The \textit{eigenvectors} of $\hat{O}$ are vectors that satisfy 
\begin{equation} 
\hat{O} \ket{\lambda_i} = \lambda_i \ket{\lambda_i}
\end{equation}
and they form an orthonormal basis, and therefore
\begin{equation} 
\braket{\lambda_i|\lambda_j} = \delta_{ij} \, .
\end{equation}
Another important fact is that
\begin{equation} 
\sum_i \ket{\lambda_i} \bra{\lambda_i} = \Id \, ,
\end{equation}
where $\ket{\lambda_i} \bra{\lambda_i}$ is the \textit{projector} operator on the vector  $\ket{\lambda_i}$. The operator $\hat{O}$ can be rewritten in terms of its eigenvectors as 
\begin{equation}\label{eq:O_as_sum_projectors} 
\sum_i \lambda_i \ket{\lambda_i} \bra{\lambda_i} = \hat{O} \, ,
\end{equation}
that means $\hat{O}$ is the weighted sum of the projectors over its eigenvectors. \\ 

\subsection{Measurements}
The \textit{measurement postulate} of quantum mechanics concerns what happens when an external observer interacts with the quantum system to measure a physical quantity of interest, represented by the Hermitian operator $\hat{O}$. It states that the system wave function \textit{collapse} in one of the eigenstates of $\hat{O}$ and the outcome of the measurement is the corresponding eigenvalue. The collapse to $\ket{\lambda_i}$ is not deterministic, but stochastic, and happens with a certain probability that is given by   
\begin{equation} 
p(\lambda_i) = |\braket{\lambda_i|\psi}|^2 \, .
\end{equation}
If one repeat the measurement experiment many times, the average value of the outcome will converge to 
\begin{equation} 
\sum_i \lambda_i \, p(\lambda_i) = \sum_i \lambda_i \braket{\psi|\lambda_i} \braket{\lambda_i|\psi} = \braket{\psi|\hat{O}|\psi}  \, ,
\end{equation}
where we used Eq.\ref{eq:O_as_sum_projectors}. We find that $\braket{\psi|\hat{O}|\psi}$, commonly referred to as the \textit{expectation value}, represents the average outcome of a measurement of the observable $\hat{O}$.

\section{Quantum Hamiltonians}

The most important operator of any physical system is the Hamiltonain operator $\hat{H}$, i.e.\ the operator that represent the quantum analogue of the energy for a classical system. The eigenvalues $\epsilon_i$ of $\hat{H}$ are the quantum \textit{energy levels} and all together they form the \textit{energy spectrum} of the system. Eigenstates with same energy are called \textit{degenerate}. The eigenstate with lower energy is often dubbed \textit{ground state}. The Hamiltonian is important because it determines how the wave function evolves with time $t$, through the famous Schroedinger equation:
\begin{equation} 
\frac{\partial}{\partial t} \ket{\psi(t)} = -i \hat{H} \ket{\psi(t)} \,. 
\end{equation}
Notice that we set the reduced Planck constant $\hbar$ to $1$ (we will always use this convention). The fact that the Hamiltonian is Hermitian, $\hat{H} = \hat{H}^{\dag}$, implies that
\begin{equation}
\frac{d}{dt} \braket{\phi(t)|\psi(t)} = -i \langle \phi(t)|\hat{H}|\psi(t) \rangle + i \langle \phi(t)|\hat{H}^{\dag}|\psi(t) \rangle = 0 \, 
\end{equation}
which implies at all time $t$
\begin{equation}
    \braket{\phi(t)|\psi(t)} = \braket{\phi(0)|\psi(0)} \, .
\end{equation}
Last equation signifies that the time evolution preserves the inner products between states. Thus, Schr\"odinger's time evolution is governed by a unitary operator $\hat{U}(t)$, such that
\begin{equation}
   \ket{\psi(t)} = \hat{U}(t) \ket{\psi(0)} \, .
\end{equation}
$\hat{U}(t)$ can be expressed explicitly as the exponential of an operator (or, equivalently, the exponential of a matrix). This is defined by using the Taylor series
\begin{equation} 
\exp( \hat{M} ) \equiv \sum_{n=0}^{\infty} \frac{\hat{M}^n}{n!} = \mathbb{1} + \hat{M} + \frac{\hat{M}^2}{2} + \frac{\hat{M}^3}{6} + ... \, .
\end{equation}
With this definition, the Schr\"odinger time evolution operator is
\begin{equation} 
\hat{U}(t) = \exp(-i \hat{H} t)  \, .
\end{equation}

\section{Pauli matrices and spin algebra}
For spin system, a very important set of observables is given by the following Pauli matrices  
\begin{equation}
\paulix = \left(
\begin{array}{cc}
 0 & 1 \\
 1 & 0 \\
\end{array}
\right) \quad \pauliy = \left(
\begin{array}{cc}
 0 & -i \\
 i & 0 \\
\end{array}
\right) \quad \pauliz = \left(
\begin{array}{cc}
 1 & 0 \\
 0 & -1 \\
\end{array}
\right) 
\end{equation}
which are interpreted as the spin magnetization along three axis $x,y,z$. Notice that we represented these operators in our standard basis $\ket{\uparrow}$, $\ket{\downarrow}$. Notice also that $\pauliz$ is diagonal in this basis, meaning that $\ket{\uparrow}$, $\ket{\downarrow}$ are eigenstates of $\pauliz$, with eigenvalues $\pm 1$:
\begin{align}
 \begin{split}
    \pauliz \ket{\uparrow} = + \ket{\uparrow} \quad  \pauliz \ket{\downarrow} = - \ket{\downarrow}  \, .
  \end{split}
\end{align}
It is very easy to show that also the other Pauli matrices have eigenvalues $\pm 1$. The corresponding eigenvectors are
\begin{equation}\label{eq:eigenstates_x}
\frac{\ket{\uparrow} + \ket{\downarrow}}{\sqrt{2}} = \frac{1}{\sqrt{2}} \left(
\begin{array}{c}
1 \\
1 \\
\end{array}
\right)
\qquad \quad
\frac{\ket{\uparrow} - \ket{\downarrow}}{\sqrt{2}} = \frac{1}{\sqrt{2}} \left(
\begin{array}{c}
1 \\
-1 \\
\end{array}
\right)
\end{equation}
for $\paulix$, and
\begin{equation}
\frac{\ket{\uparrow} + i \ket{\downarrow}}{\sqrt{2}} = \frac{1}{\sqrt{2}} \left(
\begin{array}{c}
1 \\
i \\
\end{array}
\right)
\qquad \quad
\frac{\ket{\uparrow} - i \ket{\downarrow}}{\sqrt{2}} = \frac{1}{\sqrt{2}} \left(
\begin{array}{c}
1 \\
-i \\
\end{array}
\right)
\end{equation}
for $\pauliy$. Pauli matrices are often collected together in a single vector of operators~\footnote{To be precise, we should use the notation $\vec{\hat{\sigma}}$, as the entries of the vector are operators. However, for simplicity, we will use the symbol $\vec{\sigma}$.}
\begin{equation}
    \vec{\sigma} = \left(
\begin{array}{c}
\paulix \\
\pauliy \\
\pauliz \\
\end{array}
\right) \, .
\end{equation}
We might also use the notation $\hat{\sigma}^{\alpha}$, with index $\alpha \in \{x,y,z\}$ or $\alpha \in \{1,2,3\}$ to denote the Pauli. Together with the identity matrix 
\begin{equation}
\Id = \left(
\begin{array}{cc}
 1 & 0 \\
 0 & 1 \\
\end{array}
\right) = \hat{\sigma}^0
\end{equation}
the Pauli matrices form a basis for the real vector space of $2 \times 2$ Hermitian matrices. This means that any observable can be written as a linear combination of Pauli matrices.

The mathematical importance of Pauli matrices come from the fact that they form a representation of the group SU(2), which is the group of unitary $2 \times 2$ matrices with determinant $+1$. 
We will not delve into the details of this topic, but it is sufficient to remember that if we define the \textit{spin matrices} as
\begin{equation}
    \Sop^{\alpha} = \frac{\hat{\sigma}^{\alpha}}{2} \qquad  \alpha \in \{x,y,z\} \, ,
\end{equation}
they satisfy the following very important \textit{commutation relations}
\begin{equation}
    [\Sop^{x}, \Sop^{y}] = i \Sop^{z} \qquad [\Sop^{x}, \Sop^{z}] = - i \Sop^{y} \qquad [\Sop^{y}, \Sop^{z}] = i \Sop^{x} \, .
\end{equation}
Essentially these constitute the spin analog of the canonical commutation relation between momentum $\hat{p}$ and position $\hat{q}$, namely $[\hat{q}, \hat{p}] = i$. 
From these commutation relations it also follows that the unitary operator
\begin{equation}
    \hat{U}(\theta) = \exp(i \theta \Sop^{\alpha}) \,,
\end{equation}
can be interpreted as a rotation of the spin of an angle $\theta$ around the axis $\alpha$.

\chapter{Third lecture}
\section{Combining two quantum spins}
\subsection{Tensor product of states}
Our analysis of systems of interacting quantum spins begins with a simple case in which the system consists of only two spins ($N=2$). As we discussed in the first lecture, from a classical point of view we have the following four possible states: 
\begin{equation}\label{eq:basis2}
    \{ \ket{\uparrow \uparrow} , \, \ket{\uparrow \downarrow} , \, \ket{\downarrow \uparrow} , \, \ket{\downarrow \downarrow} \} \,.
\end{equation}
However in the quantum case complex superposition are allowed, and therefore the most general wave function we should consider is 
\begin{equation}
    \ket{\psi} = \psi_0 \ket{\uparrow \uparrow} + \psi_1 \ket{\uparrow \downarrow} + \psi_2 \ket{\downarrow \uparrow} + \psi_3 \ket{\downarrow \downarrow} = \left(
\begin{array}{c}
\psi_0 \\
\psi_1 \\
\psi_2 \\
\psi_3 \\
\end{array}
\right)
\end{equation}
with $\psi_0, \psi_1, \psi_2, \psi_3$ complex numbers. Physical states satisfy the normalization condition
\begin{equation}
||\ket{\psi}||^2=|\psi_0|^2 + |\psi_1|^2 + |\psi_2|^2 + |\psi_3|^2 = 1 \, . 
\end{equation}
Our basis states can also be represented in other entirely equivalent ways, for example: 
\begin{equation}
\ket{\uparrow \downarrow} = \ket{\uparrow} \ket{\downarrow} = \ket{\uparrow} \otimes \ket{\downarrow} \, . 
\end{equation}
In the last equality, we introduced the symbol $\otimes$, representing the \textit{tensor product} between two states belonging to different Hilbert spaces. The tensor product is the mathematical construct that describes how two or more quantum systems (i.e., Hilbert spaces) combine. Naturally, we can define the tensor product of any two quantum states. For instance, if we have 
\begin{equation}
    \ket{\phi} = \left(
\begin{array}{c}
\phi_0 \\
\phi_1 \\
\end{array}
\right) \qquad  \ket{\varphi} = \left(
\begin{array}{c}
\varphi_0 \\
\varphi_1 \\
\end{array}
\right)
\end{equation}
their tensor product can be defined as follows 
\begin{align}
    \begin{split}
      \ket{\phi} \otimes \ket{\varphi} &= \left( \phi_0 \ket{\uparrow} + \phi_1 \ket{\downarrow} \right) \otimes \left( \varphi_0 \ket{\uparrow} + \varphi_1 \ket{\downarrow} \right) = \\ &=\phi_0 \varphi_0 \ket{\uparrow \uparrow} + \phi_0 \varphi_1 \ket{\uparrow \downarrow} + \phi_1 \varphi_0 \ket{\downarrow \uparrow} + \phi_1 \varphi_1 \ket{\downarrow \downarrow} \, ,  
    \end{split}
\end{align}
where we used the linearity to decompose the product of sums into a sum of products. As a vector, the tensor product $\ket{\phi} \otimes \ket{\varphi}$ is therefore can be represented as
\begin{align}
    \begin{split}
      \ket{\phi} \otimes \ket{\varphi} = \left(
\begin{array}{c}
\phi_0 \varphi_0 \\
\phi_0 \varphi_1 \\
\phi_1 \varphi_0 \\
\phi_1 \varphi_1 \\
\end{array}
\right) \, .
    \end{split}
\end{align}

\subsection{Tensor product of operators}
Now, how to define the tensor product of operators acting on different Hilbert spaces?
We can define the action of the tensor product of two operators $\hat{O}, \hat{Q}$ acting on a  tensor product of states, as follows
\begin{equation}
    (\hat{O} \otimes \hat{Q}) (\ket{\phi} \otimes \ket{\varphi}) \equiv  (\hat{O} \ket{\phi}) \otimes (\hat{Q} \ket{\varphi}) \, .
\end{equation}
For example, let us consider the operator $\pauliz \otimes \paulix$ and let us see how it acts on our two qubits basis:
\begin{align}
 \begin{split}
    (\pauliz \otimes \paulix) \ket{\uparrow \uparrow} &= (\pauliz \ket{\uparrow}) \otimes (\paulix \ket{\uparrow}) = \ket{\uparrow}\ket{\downarrow} \\
    (\pauliz \otimes \paulix) \ket{\uparrow \downarrow} &= (\pauliz \ket{\uparrow}) \otimes (\paulix \ket{\downarrow}) = \ket{\uparrow}\ket{\uparrow} \\
    (\pauliz \otimes \paulix) \ket{\downarrow \uparrow} &= (\pauliz \ket{\downarrow}) \otimes (\paulix \ket{\uparrow}) = -  \ket{\downarrow}\ket{\downarrow} \\
    (\pauliz \otimes \paulix) \ket{\downarrow \downarrow} &= (\pauliz \ket{\downarrow}) \otimes (\paulix \ket{\downarrow}) = -  \ket{\downarrow}\ket{\uparrow} \\
  \end{split}
\end{align}
That means the matrix form for $\pauliz \otimes \paulix$ is: 
\begin{equation}
\pauliz \otimes \paulix =
\left(
\begin{array}{cccc}
 0 & 1 & 0 & 0 \\
 1 & 0 & 0 & 0 \\
 0 & 0 & 0 & -1 \\
 0 & 0 & -1 & 0 \\
\end{array}
\right)
\end{equation}
Notice the structure of this matrix: in the top left, we have a $2 \times 2$ block corresponding to the matrix $\paulix$, and in the bottom right, a $2 \times 2$ block corresponding to the matrix $-\paulix$. Thus, we effectively combined $\paulix$ and $\pauliz$ by embedding $\paulix$ within the matrix elements of $\pauliz$ (which are $+1$ and $-1$ on the diagonal). Specifically, we place $\paulix$, multiplied by the factors $+1$ and $-1$, along the diagonal.

In general if our operators are represented as matrices as follows
\begin{equation}
  \hat{O} = \left(
\begin{array}{cc}
O_{00} & O_{01} \\
O_{10} & O_{11} \\
\end{array}
\right) \qquad \hat{Q} = \left(
\begin{array}{cc}
Q_{00} & Q_{01} \\
Q_{10} & Q_{11} \\
\end{array}
\right)
\end{equation}
their tensor product is:

\begin{equation}\label{eq:tens_product_op}
\hat{O} \otimes \hat{Q} =
    \left(
\begin{array}{cccc}
 O_{00} Q_{00} & O_{00} Q_{01} & O_{01} Q_{00} & O_{01} Q_{01} \\
 O_{00} Q_{10} & O_{00} Q_{11} & O_{01} Q_{10} & O_{01} Q_{11} \\
 O_{10} Q_{00} & O_{10} Q_{01} & O_{11} Q_{00} & O_{11} Q_{01} \\
 O_{10} Q_{10} & O_{10} Q_{11} & O_{11} Q_{10} & O_{11} Q_{11} \\
\end{array}
\right)
\end{equation}

\section{Two spins interacting: the XXZ model}
At this point, we want to consider some concrete physical cases, and for this reason, we introduce the following Hamiltonian
\begin{equation}
    \hat{H} = - J \left( \Sop^x_1 \Sop^x_2 + \Sop^y_1 \Sop^y_2  + \Delta \Sop^z_1 \Sop^z_2 \right) \, ,
\end{equation}
with $J>0$. Notice that, for instance, $\Sop^x_1 \Sop^x_2$ is shorthand for $\Sop^x_1 \otimes \Sop^x_2$, meaning $\Sop^x$ acts on spin 1 and $\Sop^x$ acts on spin 2. This Hamiltonian is usually dubbed XXZ model since couplings in the X and Y directions are the same, while there is an anisotropy $\Delta$ in the Z direction. Let us write the matrix form of $\hat{H}$ explicitly. 

\subsection{Exact Diagonalization for $\Delta \rightarrow \infty$}
We start by considering the case $\Delta \gg 1$ (or $\Delta \rightarrow \infty$). In this limit the Hamiltonian can be approximated as $\hat{H} = - J' \pauliz_1 \pauliz_2$, where $J' = \frac{J \Delta}{4}$. We obtain therefore the Ising model. Let us write explicitly the Hamiltonian operator $\hat{H}$ in the basis given in Eq.\ref{eq:basis2}. First we have to understand how the operator $\pauliz_1 \pauliz_2$ acts on this basis. We have:
\begin{align}
 \begin{split}
    \pauliz_1 \pauliz_2 \ket{\uparrow \uparrow} &= (+1) \ket{\uparrow \uparrow} \\
    \pauliz_1 \pauliz_2 \ket{\uparrow \downarrow} &= (-1)\ket{\uparrow \downarrow} \\
    \pauliz_1 \pauliz_2 \ket{\downarrow \uparrow} &= (-1) \ket{\downarrow \uparrow} \\
    \pauliz_1 \pauliz_2 \ket{\downarrow \downarrow} &= (+1)\ket{\downarrow \downarrow} \\
  \end{split}
\end{align}
That means the matrix form for $\hat{H}$ is: 
\begin{equation}
\hat{H} \simeq - J' 
\left(
\begin{array}{cccc}
 1 & 0 & 0 & 0 \\
 0 & -1 & 0 & 0 \\
 0 & 0 & -1 & 0 \\
 0 & 0 & 0 & 1 \\
\end{array}
\right) \, ,
\end{equation}
as we could have also directly derived using Eq.\ref{eq:tens_product_op} with operators $\hat{O}=\pauliz$ and $\hat{Q}=\pauliz$. As we see, this Ising Hamiltonian is already diagonal and therefore we do not need any diagonalization procedure: eigenvalues can be directly read on the diagonal. There are two degenerate ground states in which spins are aligned ($\ket{\uparrow \uparrow}$ and $\ket{\downarrow \downarrow}$) and two degenerates excited states in which spins are unaligned ($\ket{\uparrow \downarrow}$ and $\ket{\downarrow \uparrow}$). \\

\subsection{Exact Diagonalization for finite $\Delta$}

\begin{figure}\label{fig:energy_level}
    \centering
    \begin{tikzpicture}
        \begin{axis}[
            xlabel={$\Delta$},
            ylabel={$\epsilon/J$},
            axis lines=middle,
            width=10cm, 
            height=6cm, 
            legend style={at={(1.05,1)}, anchor=north west} 
        ]
            \addplot[
                domain=-3:3, 
                samples=100, 
                color=blue,
            ]
            {-(2-x)/4};
            \addlegendentry{$(\Delta - 2)/4$}
            
            \addplot[
                domain=-3:3, 
                samples=100, 
                color=red,
            ]
            {-(-2-x)/4};
            \addlegendentry{$(\Delta+2)/4$}
            
            \addplot[
                domain=-3:3, 
                samples=100, 
                color=violet,
            ]
            {-x/4};
            \addlegendentry{$-\Delta/4$}
        \end{axis}
    \end{tikzpicture}
    \caption{Spectrum of the XXZ Hamiltonian for $N=2$ spins as a function of the anisotropy $\Delta$.}
\end{figure}
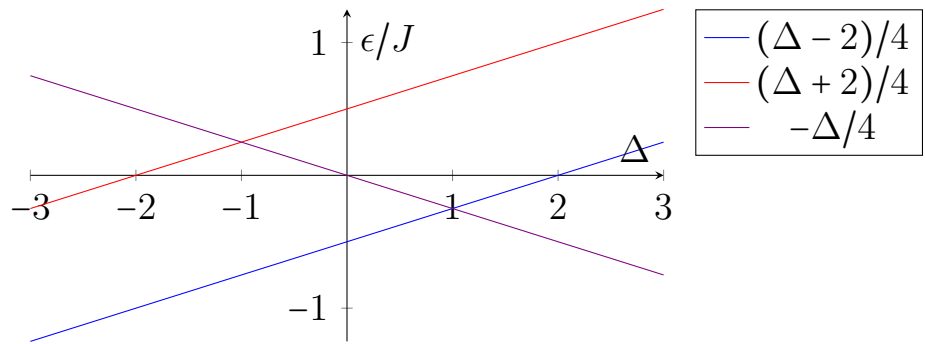

Now we move to the general case of finite anisotropy $\Delta$. We rewrite the Hamiltonian in the following way:
\begin{align}
 \begin{split}
     \hat{H} = - \frac{J}{4} \left( \paulix_1 \paulix_2 + \pauliy_1 \pauliy_2  + \Delta \pauliz_1 \pauliz_2 \right) = - J \left( \frac{1}{2} \left( \pauliplus_1 \pauliminus_2 + \pauliminus_1 \pauliplus_2 \right) + \frac{\Delta}{4} \pauliz_1 \pauliz_2 \right) \, .
 \end{split}
\end{align}
In last equality we introduced the following ladder operators
\begin{equation}
\pauliplus = \frac{\paulix + i \pauliy}{2} = \left(
\begin{array}{cc}
 0 & 1 \\
 0 & 0 \\
\end{array}
\right) 
\qquad \pauliminus = \frac{\paulix - i \pauliy}{2} = 
\left(
\begin{array}{cc}
 0 & 0 \\
 1 & 0 \\
\end{array}
\right)
\end{equation}
which satisfied this useful property: 
\begin{align}
 \begin{split}
   \pauliplus \ket{\downarrow} = \ket{\uparrow} \qquad    \pauliplus \ket{\uparrow} &= 0 \\ 
   \pauliminus \ket{\downarrow} = 0 \qquad    \pauliminus \ket{\uparrow} &= \ket{\downarrow} \, .
  \end{split}
\end{align}
Now we can easily understand how the operator $\pauliplus_1 \pauliminus_2$ acts on our basis. We have:
\begin{align}
 \begin{split}
    \pauliplus_1 \pauliminus_2 \ket{\uparrow \uparrow} &= 0 \\
    \pauliplus_1 \pauliminus_2 \ket{\uparrow \downarrow} &= 0 \\
    \pauliplus_1 \pauliminus_2 \ket{\downarrow \uparrow} &= \ket{\uparrow \downarrow} \\
    \pauliplus_1 \pauliminus_2 \ket{\downarrow \downarrow} &= 0 \\
  \end{split}
\end{align}
and therefore 
\begin{align}
\pauliplus_1 \pauliminus_2 = 
\left(
\begin{array}{cccc}
 0 & 0 & 0 & 0 \\
 0 & 0 & 1 & 0 \\
 0 & 0 & 0 & 0 \\
 0 & 0 & 0 & 0 \\
\end{array}
\right)
\end{align}
Similiarly one can write down $\pauliminus_1 \pauliplus_2$. Summing together these matrices, we finally get the expression for the Hamiltonian, that is
\begin{equation}
\hat{H} = - \frac{J}{4}
\left(
\begin{array}{cccc}
 \Delta & 0 & 0 & 0 \\
 0 & -\Delta & 2 & 0 \\
 0 & 2 & -\Delta & 0 \\
 0 & 0 & 0 & \Delta \\
\end{array}
\right)
\end{equation}
We want to diagonalize $\hat{H}$ in order to find the energy spectrum and the ground state. First we notice that part of the matrix is already diagonal in the basis we use. Indeed the vectors
\begin{equation}
\ket{\uparrow \uparrow} = \left(
\begin{array}{c}
1 \\
0 \\
0 \\
0 \\
\end{array}
\right) \qquad 
\ket{\downarrow \downarrow} = \left(
\begin{array}{c}
0 \\
0 \\
0 \\
1 \\
\end{array}
\right)
\end{equation}
are degenerate eigenvectors with energy $-\frac{J \Delta}{4}$. We have now to diagonalize only the central $2 \times 2$ sub-matrix of $\hat{H}$, that we name $\tilde{H}$:
\begin{equation}
\tilde{H} = \left(
\begin{array}{cc}
-\Delta & 2 \\
2 & -\Delta \\
\end{array}
\right)
\end{equation}
In order to find the eigenvalues of $\tilde{H}$ we have to find its \textit{characteristic polynomial} \cite{enwiki:1220838427}, that is
\begin{equation}\label{eq:carpol}
P(\lambda)= \det(\tilde{H} - \lambda \mathbb{1}) = 
\det \left(
\begin{array}{cc}
-\Delta-\lambda & 2 \\
2 & -\Delta-\lambda \\
\end{array}
\right) = (\Delta+\lambda)^2 - 4
\end{equation}
where $\lambda$ is the variable of the polynomial $P(\lambda)$. Now we have to determine the roots of the characteristic polynomial, i.e.\ the solutions of the equation $P(\lambda)=0$. In our case, we get a simple quadratic equation, from which we obtain 
\begin{equation}
\Delta +\lambda=\pm 2 \, .
\end{equation}
Therefore, we find two distinct solutions:
\begin{equation}
\lambda = 2 - \Delta \qquad  \lambda = -2-\Delta \, .
\end{equation}
To find the corresponding eigenvectors we have to replace these values in $\tilde{H} - \lambda \mathbb{1}$, obtaining the two matrices
\begin{equation}
\tilde{H} - (2 - \Delta) \cdot \mathbb{1} = \left(
\begin{array}{cccc}
-2 & 2 \\
2 & -2 \\
\end{array}
\right)
\qquad \quad
\tilde{H} + (2+\Delta) \cdot \mathbb{1} = \left(
\begin{array}{cccc}
2 & 2 \\
2 & 2 \\
\end{array}
\right)
\end{equation}
Now, we have to find the \textit{null space} (also named \textit{kernel}) of these two matrices. The null space is the set of vectors that are mapped to $0$ by the application of a matrix. It can be determined through \textit{Gaussian elimination}. However in our simple case one can easily realize that the two vectors
\begin{equation}
\frac{1}{\sqrt{2}}\left(
\begin{array}{cccc}
1 \\
1 \\
\end{array}\right)
\qquad \quad
\frac{1}{\sqrt{2}}\left(
\begin{array}{cccc}
1 \\
-1 \\
\end{array}\right)
\end{equation}
form respectively the null space of our matrices. Now that we have found the eigenvectors of the sub-matrix $\tilde{H}$, we can go back and embed them into vectors of length $4$. We then find:
\begin{equation}
\frac{1}{\sqrt{2}}\left(
\begin{array}{cccc}
0 \\
1 \\
1 \\
0 \\
\end{array}\right) = \frac{\ket{\uparrow \downarrow} + \ket{\downarrow \uparrow}}{\sqrt{2}}
\qquad \quad
\frac{1}{\sqrt{2}}\left(
\begin{array}{cccc}
0 \\
1 \\
-1 \\
0 \\
\end{array}\right) = \frac{\ket{\uparrow \downarrow} - \ket{\downarrow \uparrow}}{\sqrt{2}}
\end{equation}
The state on the left is known as \textit{triplet}, the state on the right is the \textit{singlet}. We found the four eigenvalues of $\hat{H}$, which are
\begin{equation}\label{eq:eigenergies}
    \epsilon = -J\frac{\Delta}{4} , \, \epsilon = 
   J\frac{\Delta-2}{4}, \, \epsilon =  J\frac{\Delta+2}{4} \, ,
\end{equation}
where the first is denegerate. 
To understand who is the ground state let us plot these values as a function of the anisotropy $\Delta$. The plot is shown in Figure \ref{fig:energy_level}. 

As you can see, depending on the parameter $\Delta$, the nature of the ground state change. In particular:

\begin{itemize}
    \item for $\Delta < 1$, the triplet $\frac{\ket{\uparrow \downarrow} + \ket{\downarrow \uparrow}}{\sqrt{2}}$ is the ground state;
    \item for $\Delta = 1$, the ground state is degenerate because the triplet $\frac{\ket{\uparrow \downarrow} + \ket{\downarrow \uparrow}}{\sqrt{2}}$ and the two ferromagnetic states $\ket{\uparrow \uparrow}$, $\ket{\downarrow \downarrow}$ have the same energy;
    \item for $\Delta > 1$, the two ferromagnetic states $\ket{\uparrow \uparrow}$, $\ket{\downarrow \downarrow}$ are the ground state.
\end{itemize}

The point $\Delta=1$ is named \textit{level crossing}, since the ground state of the system change from the triplet state to the ferromagnetic state. At level crossing usually the ground state undergoes changes in nature, for instance, indicated by a sudden change in the value of a certain observable considered as an \textit{order parameter}. This is a signal of a phase transition, which is termed \textit{quantum phase transition} because it is not associated (as in the classical case) with thermal fluctuations, but with quantum fluctuations. Naturally, the actual transition is only possible in the thermodynamic limit, that is, when $N \rightarrow \infty$. A more detailed description of quantum phase transitions and level crossing will be provided in Lecture \ref{ch:cpt}.
\chapter{Fourth lecture}
\section{Curse of dimensionality}

The true complexity of quantum systems emerges when the number 
$N$ of constituents becomes large. Therefore, let's suppose we have $N\gg 1$ spins. What does the wave function describing the state look like? The classical configurations of the system are:
\begin{equation}\label{eq:basisN}
    \{ \ket{\uparrow ... \uparrow \uparrow} , \, \ket{\uparrow ... \uparrow \downarrow} , \, \ket{\uparrow ... \downarrow \uparrow} , \, ... \,  \ket{\downarrow ... \downarrow \downarrow} \} \,.
\end{equation}
and therefore the most general wave function is
\begin{equation}
    \ket{\psi} = \psi_0 \ket{\uparrow ... \uparrow \uparrow} + \psi_1 \ket{\uparrow ... \uparrow \downarrow} + \psi_2 \ket{\uparrow ... \downarrow \uparrow} + ... + \psi_{2^N - 1} \ket{\downarrow ... \downarrow \downarrow}
\end{equation}
which can also be represented as a vector as 
\begin{equation}
    \ket{\psi} = \left(
\begin{array}{c}
\braket{\uparrow ... \uparrow \uparrow| \psi} \\
\braket{\uparrow ... \uparrow \downarrow| \psi} \\
\braket{\uparrow ... \downarrow \uparrow| \psi} \\
\vdots \\
\braket{\downarrow ... \downarrow \downarrow| \psi}  \\
\end{array}
\right) = \left(
\begin{array}{c}
\psi_0 \\
\psi_1 \\
\psi_2 \\
\vdots \\
\psi_{2^N - 1} \\
\end{array}
\right)
\end{equation}
This representation offers a full description
of the wave function $\ket{\psi}$. However, it also  requires to store all the components of the vector $\ket{\psi}$, that means $2^N$ complex numbers. Thus, for $N=30$ spins we would need to store 
\begin{equation}
2^N = 1,073,741,824 \simeq 10^9  
\end{equation}
complex numbers, which requires approximately 16 GB of memory on a hard disk. For $N=50$, one needs to store
\begin{equation}
 2^N = 1,125,899,906,842,624 \simeq 10^{15}   
\end{equation}
complex numbers, that is approximately $16$PB $= 16 \cdot 10^6$GB of memory. This amount of memory is extraordinarily large, far exceeding the capacity of most current storage systems. Similarly, processing $\ket{\psi}$, for instance to compute a matrix element, requires an exponentially large number of operations. As a matter of fact, simulating quantum systems of size larger than $N \approx 30$ with exact numerical techniques is impossible, even with the largest super computers available today.

\section{Generation of the XXZ Hamiltonian with $N$ spins}
Now we will consider the Hamiltonian of the $XXZ$ model for $N$ quantum spins, that is:
\begin{align}
\begin{split}
    \hat{H} &= - J \sum_{i=0}^{N-1} \left( \Sop^x_{i} \Sop^x_{i+1} + \Sop^y_{i} \Sop^y_{i+1}  + \Delta \Sop^z_{i} \Sop^z_{i+1} \right) = \\
   &= - J \sum_{i=0}^{N-1} \left( \frac{1}{2} \left( \pauliplus_{i} \pauliminus_{i+1} + \pauliminus_{i} \pauliplus_{i+1} \right) + \frac{\Delta}{4} \pauliz_{i} \pauliz_{i+1} \right)\, \, .
\end{split}  
\end{align}
Notice that for convenience sites have been labelled $0,1,... \, N-1$ instead of $1,2,.. N$. The task is to write down a practical algorithm (or a code) to obtain the matrix representation of $\hat{H}$ as a matrix of size $2^N \times 2^N$. First, we need to implement two basic functions: 
\begin{itemize}
    \item a function $\texttt{bit}(x, N)$ that takes an integer $x \in \{0, 1, 2, \ldots, 2^N-1\}$ and returns its $N$-digits binary representation $b \in \{0, 1\}^N$ as an array of length $N$;
    \item a function $\texttt{num}(b, N)$ that takes an array $b$ representing $N$ binary digits and returns the corresponding integer $x$.
\end{itemize}
We will have for instance
\begin{equation}\label{eq:bin}
    \texttt{bit}(10,4) = (1,0,1,0)
\end{equation}
and 
\begin{equation}\label{eq:num}
\texttt{num}\big((1,1,0,1),4\big) = 13 \, .
\end{equation}
We also need a function $\texttt{flip}(b, i, j)$ that takes a binary array $b$ of $N$ digits and returns an array of the same length in which digits $i$ and $j$ (with $i, j \in \{0, 1, \ldots, N-1\}$) have been flipped (i.e., $0 \rightarrow 1$, $1 \rightarrow 0$). For instance
\begin{equation}\label{eq:flip}
\texttt{flip}\big((1,1,0,1),0,3\big) = (0,1,0,0) \, .
\end{equation}
Now the idea is to prepare $H$ as a matrix full of zeros and then fill the entries (matrix elements) which are non zero. To do so, we do a for loop over all the states of the computational basis, which are labelled by a single integer $x \in \{0, 1, 2, \ldots, 2^N-1\}$ and converted to strings by $b = \texttt{bit}(x,N)$. Basically, we have just to see how the three operators $\frac{1}{2} \pauliplus_{i} \pauliminus_{i+1}$, $\frac{1}{2} \pauliminus_{i} \pauliplus_{i+1}$, $\frac{\Delta}{4} \pauliz_{i} \pauliz_{i+1}$ act on this basis. To this end, we note that:
\begin{itemize}
    \item the $\pauliz_{i} \pauliz_{i+1}$ operator simply contributes with $+1$ when two spins are aligned (i.e.\ digits $i$ and $i+1$ are both $0$ or both $1$), and with $-1$ when two spins are anti-aligned (digits $i$ and $i+1$ are $0$ and the other $1$);
    \item the operator $\sigma^{+}_{i} \sigma^{-}_{i+1}$ acts non-trivially only if bits $i$ and $i+1$ are $1$ and $0$, respectively, and in this case, it flips them to $0$ and $1$. Similarly, the operator $\sigma^{-}_{i} \sigma^{+}_{i+1}$ acts non-trivially only if bits $i$ and $i+1$ are $0$ and $1$, respectively, and in this case, it flips them to $1$ and $0$.
\end{itemize}

Putting these observations together, we can outline the Algorithm \ref{alg:XXZ_Ham}.

\begin{algorithm}[h!]
\caption{Generation of the $XXZ$ Hamiltonian}\label{alg:XXZ_Ham}
\begin{flushleft}
\hspace*{\algorithmicindent} \textbf{Input}: the system size $N$ and the parameters $J, \Delta$
\end{flushleft}
\begin{algorithmic}[1]
\State Initialize $H$ as a $2^N \times 2^N$ matrix with all entries $0$ (or a sparse matrix)
\For{($x=0$, $x=2^N-1$, $x++$)}
     \State Set $b$ to be the binary form of $x$ with $N$ digits: $b = \texttt{bit}(x,N)$
     \For{($i=0$, $i=N-1$, $i++$)}
           \State Set $j = \mod(i+1, N)$
           \If{$b[i]=b[j]$} 
               \State Do $H[x,x] = H[x,x] - \frac{J \Delta}{4}$
            \Else
               \State Do $H[x,x] = H[x,x] + \frac{J \Delta}{4}$
               \State Flip digits $i,j$ in $b$ obtaining $b' = \texttt{flip}(b,i,j)$. 
               \State Set: $y = \texttt{num}(b',N)$
               \State Do $H[x,y] = H[x,y] - \frac{J}{2}$
\EndIf 
     \EndFor
\EndFor
\end{algorithmic}
\begin{flushleft}
\hspace*{\algorithmicindent} \textbf{Output}: the $XXZ$ Hamiltonian matrix $H$
\end{flushleft}
\end{algorithm}

\section{Generation of generic spin models}\label{sec:ising_quantum}

We have examined a significant model, the XXZ model; however, there is a wide array of important spin chain Hamiltonians. For instance, in our first lesson, we considered a general model of classical Ising-like Hamiltonians. What is the quantum equivalent of this Hamiltonian? The answer is very simple and is obtained by formally applying the substitution 
\begin{equation}
    \spin_{i} \rightarrow \pauliz_{i} \, ,
\end{equation}
where on the left we have the classical spin state, and on the right the corresponding quantum operator. It is often said that classical spins are ``promoted'' to be quantum operators. We therefore get
\begin{equation}
    \hat{H} = - \frac{1}{2} \sum_{i=1}^{N} \sum_{j=1}^{N}  J_{ij} \pauliz_{i} \pauliz_{j} - \sum_{i=1}^N g_i \pauliz_{i} \, ,
\end{equation} 
However, this model is, in a certain sense, not quantum. In fact,  all the terms involved in the Hamiltonian commute with each other, as they only contain Pauli $z$ operators and $[\pauliz_{i}, \pauliz_{j}]=0$. Furthermore, all eigenstates of $\hat{H}$ are simply the basis states $\ket{\spin_1, \spin_2, \ldots, \spin_N}$, indicating that they lack a fundamental characteristic of quantum states: superposition. Therefore, we cannot expect truly quantum features from this model. For this reason, an additional term involving the $\paulix_i$ operators is usually added
\begin{equation}
    \hat{H} = - \frac{1}{2} \sum_{i=1}^{N} \sum_{j=1}^{N}  J_{ij} \pauliz_{i} \pauliz_{j} - \sum_{i=1}^N g_i \pauliz_{i}  - \sum_{i=1}^N h_i \paulix_{i} \, .
\end{equation}
This Hamiltonian represents the \textit{transverse field Ising model}, which is one of the paradigmatic quantum models for spin chains, analogous to the standard Ising model in the classical case. Typically, interactions are considered only between nearest neighbors, and the magnetic fields are assumed to be uniform across all sites. In the case of a one-dimensional lattice, this means:
\begin{equation}\label{eq:quantum_ising}
    \hat{H} = - J \sum_{i=1}^{N-1} \pauliz_{i} \pauliz_{i+1} - g \sum_{i=1}^N \pauliz_{i} - h \sum_{i=1}^N \paulix_{i} \, .
\end{equation}
The transverse field $h$ gives rise to non-trivial \textit{quantum fluctuations}. It can be seen that there is a precise relationship between these quantum fluctuations and the thermal fluctuations of the classical model. \\

For all these reasons, it may be useful to have routines capable of generating any spin model, as for instance Eq.\ref{eq:quantum_ising}, not just the XXZ model. Therefore, we need to adapt what we have done to be more generalizable. First, we will define functions that implement how the Pauli operators $x,y,z$ at a certain site $i$ of the chain act on the elements of the computational basis we are using. As we have seen, it is convenient to identify these elements with a single integer $x$ ranging from $0$ to $2^N-1$. We must first note that every time a Pauli operator, for example $\hat{\sigma}^{\alpha}_i$, is applied to the state $\ket{x}$, a new state of the computational basis, which we can call $\ket{y}$, is always obtained, except for a multiplicative coefficient. In formula: $\hat{\sigma}^{\alpha}_i \ket{x} = c \ket{y}$. For instance, with $N=3$ spins, 
\begin{equation}
\hat{\sigma}^{x}_{1} \ket{5} = \hat{\sigma}^{x}_{1} \ket{101} = \hat{\sigma}^{x}_{1} \ket{\downarrow \uparrow \downarrow} = \ket{\downarrow \downarrow \downarrow} = \ket{1 1 1} = \ket{7} \, , 
\end{equation}
where we freely switched between different ways of labeling the basis states (as integers, as strings of $0,1$, or as strings of $\uparrow,\downarrow$). In this case we have therefore $y=5$ and $c=1$. Notice that $\paulix_i$ (and also $\pauliy_i$) acts by flipping a single bit, so we need to extend our $\texttt{flip}$ function to the case we just want to flip one bit. Thus we will have for instance 
\begin{equation}
    \texttt{flip}\big((1,0,1),1\big) = (1,1,1) \, .
\end{equation}
We are now ready to write the functions that implement $\paulix_i, \pauliy_i, \pauliz_i$ on a given state $\ket{x}$ (see Algorithms \ref{alg:X_i}, \ref{alg:Y_i}, \ref{alg:Z_i}).

\begin{algorithm}[h!]
\caption{Action of Pauli operator $\paulix_i$ on basis state $x$}\label{alg:X_i}
\begin{flushleft}
\hspace*{\algorithmicindent} \textbf{Input}: system size $N$, site index $i$, an integer $x$ labeling an element of the computational basis
\end{flushleft}
\begin{algorithmic}[1]
\State Set $b$ to be the binary form of $x$ with $N$ digits: $b = \texttt{bit}(x,N)$
\State Create a new string $b'$ by flipping bit $i$ of $b$:  $b' = \texttt{flip}(b,i)$.
\State Set $y = \texttt{num}(b',N)$
\State Set $c=1$
\end{algorithmic}
\begin{flushleft}
\hspace*{\algorithmicindent} \textbf{Output}: the integer $y$ labeling the new basis state, the coefficient $c$
\end{flushleft}
\end{algorithm}

\begin{algorithm}[h!]
\caption{Action of Pauli operator $\pauliy_i$ on basis state $x$}\label{alg:Y_i}
\begin{flushleft}
\hspace*{\algorithmicindent} \textbf{Input}: system size $N$, site index $i$, an integer $x$ labeling an element of the computational basis
\end{flushleft}
\begin{algorithmic}[1]
\State Set $b$ to be the binary form of $x$ with $N$ digits: $b = \texttt{bit}(x,N)$
\State Create a new string $b'$ by flipping bit $i$ of $b$:  $b' = \texttt{flip}(b,i)$.
\State Set $y = \texttt{num}(b',N)$
\If{$b[i]=0$} 
               \State Set $c=i$
\Else
               \State Set $c=-i$
\EndIf 
\end{algorithmic}
\begin{flushleft}
\hspace*{\algorithmicindent} \textbf{Output}: the integer $y$ labeling the new basis state, the coefficient $c$
\end{flushleft}
\end{algorithm}

\begin{algorithm}[h!]
\caption{Action of Pauli operator $\pauliz_i$ on basis state $x$}\label{alg:Z_i}
\begin{flushleft}
\hspace*{\algorithmicindent} \textbf{Input}: system size $N$, site index $i$, an integer $x$ labeling an element of the computational basis
\end{flushleft}
\begin{algorithmic}[1]
\State Set $y = x$
\State Set $b$ to be the binary form of $x$ with $N$ digits: $b = \texttt{bit}(x,N)$
\If{$b[i]=0$} 
               \State Set $c=1$
\Else
               \State Set $c=-1$
\EndIf 
\end{algorithmic}
\begin{flushleft}
\hspace*{\algorithmicindent} \textbf{Output}: the integer $y$ labeling the new basis state, the coefficient $c$
\end{flushleft}
\end{algorithm}

With the same tools, one can also write coupling operators like $\paulix_i \paulix_j$. This will be constructed as in Algorithm \ref{alg:XX_i}.

\begin{algorithm}[h!]
\caption{Action of Pauli operator $\paulix_i \paulix_j$ on basis state $x$}\label{alg:XX_i}
\begin{flushleft}
\hspace*{\algorithmicindent} \textbf{Input}: system size $N$, site indices $i,j$, an integer $x$ labeling an element of the computational basis
\end{flushleft}
\begin{algorithmic}[1]
\State Set $b$ to be the binary form of $x$ with $N$ digits: $b = \texttt{bit}(x,N)$
\State Create a new string $b'$ by flipping bit $i$ and $j$ of $b$:  $b' = \texttt{flip}(b,i,j)$.
\State Set $y = \texttt{num}(b',N)$
\State Set $c=1$
\end{algorithmic}
\begin{flushleft}
\hspace*{\algorithmicindent} \textbf{Output}: the integer $y$ labeling the new basis state, the coefficient $c$
\end{flushleft}
\end{algorithm}

With these functions, or others very similar, any spin Hamiltonian should be easily generable. In Chapter \ref{ch:code}, we provide a working Python implementation of all these functions.

\section{Spin models experimentally}

\begin{figure}[t!]
\centering
\includegraphics[width=0.75\linewidth]{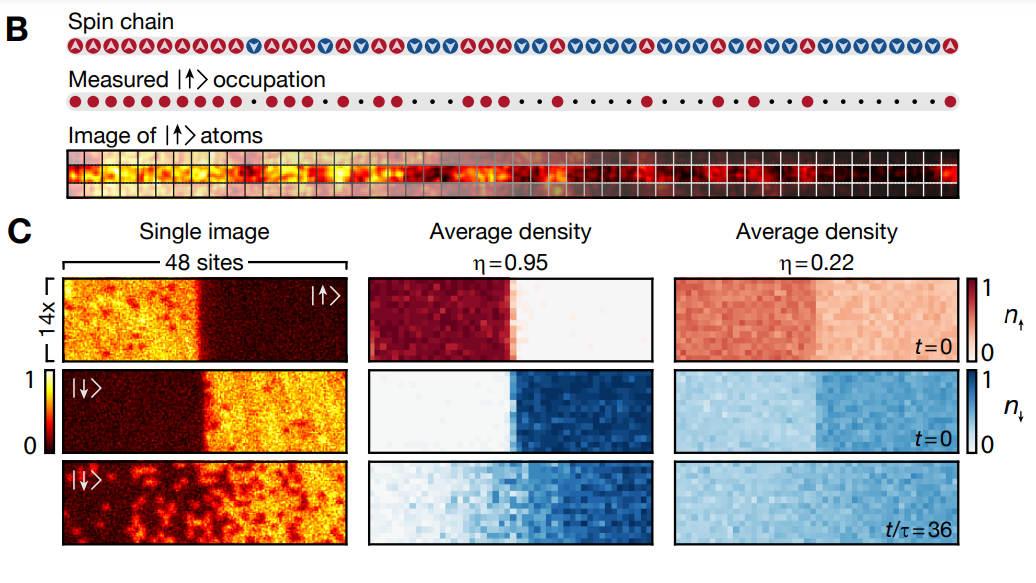}
\caption{The XXZ can be realized experimentally! Figure adapted from Ref.\cite{Wei_2022}.} 
\label{fig:bloch}
\end{figure}

While the models we’ve discussed may initially appear as theoretical toy models with limited practical applications, this is not the case. In fact, they can be experimentally implemented. Immanuel Bloch's group obtained the XXZ model using bosonic 87Rb atoms trapped in an optical lattice (see Fig.\ref{fig:bloch}). In this setup, the two spin states, $\ket{\uparrow}$ and $\ket{\downarrow}$, are encoded in the two hyperfine ground states of the atoms. The system is described by the Bose-Hubbard Hamiltonian, which can be mapped to the XXZ model in the limit of strong interactions. The local occupation number, i.e., the operator $\Sop^z_{i}$, can be measured by fluorescence imaging.

\chapter{Fifth lecture}
\section{Symmetries: $U(1)$ symmetry}
A very important linear algebra theorem 
states that when two Hermitian operators $\hat{H}$ and $\hat{O}$ commute each other, $[\hat{H},\hat{O}]=0$, then it is possible to \textit{simultaneously} diagonalize them. This means that is possible to find a single unitary $\hat{U}$, such that both $\hat{U} \hat{H} \hat{U}^{\dag}$ and $\hat{U} \hat{O} \hat{U}^{\dag}$ are diagonal. This property often allow to greatly simplify the problem of finding the spectrum of an Hamiltonian. 

To see a practical example, let us consider again the XXZ model for $N=2$ spins:
\begin{equation}
    \hat{H} = - J \left( \Sop^x_1 \Sop^x_2 + \Sop^y_1 \Sop^y_2  + \Delta \Sop^z_1 \Sop^z_2 \right) \, . 
\end{equation}
Let us consider now the following operator
\begin{equation}
    \hat{S}^z = \sum_{i=1}^N \Sop^z_i = \Sop^z_1 + \Sop^z_2 
\end{equation}
which physically represent the total magnetization along the $z$ direction. As a matrix $\hat{S}^z$ is 
\begin{equation}
\hat{S}^z = 
\left(
\begin{array}{cccc}
 1 & 0 & 0 & 0 \\
 0 & 0 & 0 & 0 \\
 0 & 0 & 0 & 0 \\
 0 & 0 & 0 & -1 \\
\end{array}
\right) \, .
\end{equation}
Let us try to calculate the commutator between this operator and the various terms of our Hamiltonian. First, we note that $\hat{S}^z$ commutes with $\Sop^z_1 \Sop^z_2$ because the $\Sop^z$ operators commute. We proceed with the other two terms and use the spin commutation rules, and the fact that spins on different sites always commute with each other. We have 
\begin{align}
    \begin{split}
        &[\Sop^x_1 \Sop^x_2,\hat{S}^z] = \\ &= [\Sop^x_1 \Sop^x_2,\Sop^z_1] + [\Sop^x_1 \Sop^x_2,\Sop^z_2] = [\Sop^x_1,\Sop^z_1] \Sop^x_2 + \Sop^x_1 [\Sop^x_2,\Sop^z_2] = - i \Sop^y_1 \Sop^x_2 - \Sop^x_1 \Sop^y_2 
    \end{split}
\end{align}
and 
\begin{align}
    \begin{split}
        &[\Sop^y_1 \Sop^y_2,\hat{S}^z] = \\ &= [\Sop^y_1 \Sop^y_2,\Sop^z_1] + [\Sop^y_1 \Sop^y_2,\Sop^z_2] = [\Sop^y_1,\Sop^z_1] \Sop^y_2 + \Sop^y_1 [\Sop^y_2,\Sop^z_2] = + i \Sop^x_1 \Sop^y_2 + \Sop^y_1 \Sop^x_2 \, .  
    \end{split}
\end{align}
Summing together these results we obtain
\begin{equation}
    [\hat{H},\hat{S}^z] = 0 \, ,
\end{equation}
meaning that the total $z$-magnetization commute with our Hamiltonian. A similar calculation can be done in for $XXZ$ Hamiltonian with $N$ spins. 

Now, since $\hat{H}$ and $\hat{S}^z$ commute, it means that they share a common set of eigenstates. The operator $\hat{S}^z$ measures the total $z$-component of the magnetization, and its eigenstates are our standard spin basis states $\ket{\sigma_1, \sigma_2, \ldots, \sigma_N}$. Since $\hat{H}$ commutes with $\hat{S}^z$, $\hat{H}$ must preserve the eigenvalues of $\hat{S}^z$. Consequently, $\hat{H}$ does not mix states with different $M^z$ values, but only acts within subspaces of fixed $M^z$. Therefore, in the standard spin basis, $\hat{H}$ takes a block diagonal form where each block corresponds to a subspace with a fixed total magnetization $M^z$:
\footnotesize
\begin{equation}\label{eq:hamblockdiagonal}
\hat{H} = 
  \setlength{\arraycolsep}{0pt}
  \begin{pmatrix}
    \,\fbox{$\hat{S}^z=N/2$} & \, & \,  \\
    & \,\fbox{$\hat{S}^z=N/2-1$} & \, & \,  \\
    & \, & \ddots \\
    & \, &  & \fbox{\raisebox{1.5ex}[5.ex]{$\hat{S}^z=0$}} & \,  & \\
    & \, & \, & \, & \ddots \\
    \, & \, & \, & \, & \, & \, \fbox{$\hat{S}^z=-N/2+1$}\, & \,  \\
    \, & \, & \, & \, & \, & \, & \, & \,\fbox{$\hat{S}^z=-N/2$}\, \\
  \end{pmatrix}
\end{equation}
\normalsize
Notice that we have assumed $N$ to be an even number, otherwise, the block with zero magnetization would not be present. The block diagonal form leads to significant simplifications in the analysis of the system:
Indeed, instead of dealing with the full Hamiltonian matrix $\hat{H}$, which has a size of $2^N \times 2^N$, we can work with the smaller blocks. Furthermore, the independent nature of the blocks allows for parallel computation: each block can be diagonalized separately and simultaneously on different processors, making efficient use of computational resources. \\

The symmetry we just discussed, is usually named U(1) symmetry. Indeed, if we define the unitary operator
\begin{equation}
    \hat{U}(\theta) = e^{-i \theta \hat{S}^z} \, ,
\end{equation}
which physically represent the rotation of all spins around the $z$-axis, we can easily realize that 
\begin{equation}
    \hat{U}(\theta) \hat{H} \hat{U}^{\dag}(\theta) =  \hat{H} \, ,
\end{equation}
which means that the Hamiltonian is left invariant by any rotation around the $z$-axis. The group of rotations $\hat{U}(\theta)$ is abelian, because 
$[\hat{U}(\theta),\hat{U}(\theta')]=0$ for any angles $\theta, \theta'$, and it is isomorphic to U(1), the group of complex phases.

\section{Symmetries: $SU(2)$ symmetry}

Let us consider again the $XXZ$ Hamiltonian in the simple case of just $N=2$ spins. Let us also set $\Delta = 1$. In this case the Hamiltonian is:
\begin{equation}
\hat{H} = - J \left( \Sop^x_1 \Sop^x_2 + \Sop^y_1 \Sop^y_2  + \Sop^z_1 \Sop^z_2 \right)  = - J \vec{\Sop}_1 \cdot \vec{\Sop}_2 \, ,
\end{equation}
where we used the vectors of spin operators $\vec{\Sop}_1$ and $\vec{\Sop}_2$. Now we use a trick to formally rewrite $\hat{H}$ in the following way
\begin{equation}
\hat{H} = - \frac{J}{2} \left( (\vec{\Sop}_1 + \vec{\Sop}_2)^2 -\vec{\Sop}_1^2 - \vec{\Sop}_2^2 \right) \, .
\end{equation}
Let us notice that 
\begin{align}
    \begin{split}
      \vec{\Sop}_1^2 &= \vec{\Sop}_1 \cdot \vec{\Sop}_1 = (\Sop^x_1)^2 + (\Sop^y_1)^2  + (\Sop^z_1)^2 = \\  &=\frac{1}{4} \left( (\paulix_1)^2 + (\pauliy_1)^2  + (\pauliz_1)^2 \right) =  \frac{1}{4} 3 \cdot \mathbb{1} = \frac{1}{2} ( \frac{1}{2} + 1) \, .
    \end{split}
\end{align}
In fact, when for spin $S$, the total spin operator $\vec{\Sop}^2$ is proportional to the identity and takes the value $S(S+1)$, that in our case gives $\frac{1}{2} (\frac{1}{2} + 1) = \frac{3}{4}$. Thus, we can rewrite the Hamiltonian as
\begin{equation}\label{eq:hamsu2}
\hat{H} = - \frac{J}{2} \left( (\vec{\Sop}_1 + \vec{\Sop}_2)^2 - \frac{3}{2} \right) \, .
\end{equation}
Now, drawing from the theory of quantum angular momentum, we recall that when two spin $1/2$ particles are combined, their total angular momentum can only result in two possible values: either $S=1$ or $S=0$. This can be expressed symbolically as 
\begin{equation}
    \frac{1}{2} \otimes \frac{1}{2} = 1 \oplus 0 \, ,
\end{equation}
and it arises due to the rules of angular momentum addition in quantum mechanics. The subspace with total spin $S=1$ is spanned by the three states 
\begin{equation}
    \ket{\uparrow \uparrow} \, , \frac{\ket{\uparrow \downarrow} + \ket{\downarrow \uparrow}}{\sqrt{2}} \,, \ket{\downarrow \downarrow} \, ,
\end{equation}
whereas the spin $S=0$ space is given by: 
\begin{equation} \frac{\ket{\uparrow \downarrow} - \ket{\downarrow \uparrow}}{\sqrt{2}} \, .
\end{equation}
In the first space, we will have 
\begin{equation}
(\vec{\Sop}_1 + \vec{\Sop}_2)^2 = 1 (1 + 1) = 2 \, ,
\end{equation}
while in the second 
\begin{equation}
(\vec{\Sop}_1 + \vec{\Sop}_2)^2 = 0 \, .
\end{equation}
By replacing these values into Eq.\ref{eq:hamsu2}, we get $\hat{H} = - J\frac{1}{4}$ in the spin $1$ sector, and $\hat{H} = + J \frac{3}{4}$ in the spin $0$ sector. These values are compatible with the eigenenergies we found in Eq.\ref{eq:eigenergies} (for $\Delta=1$). \\

The form of Eq.\ref{eq:hamsu2} makes it evident that in this case the Hamiltonian has another symmetry. Indeed, since $\hat{H}$ is (a part for irrelevant constants) the total spin $(\vec{\Sop})^2$ (where $\vec{\Sop} = \vec{\Sop}_1 + \vec{\Sop}_2$), then it has to commute with its $x,y,z$ components, i.e.\
\begin{equation}
    [\hat{H},\hat{S}^x] = 0 \qquad [\hat{H},\hat{S}^y] = 0 \qquad [\hat{H},\hat{S}^z] = 0 \, .
\end{equation}
This symmetry is also present in the case of $N$ spins, i.e. $\hat{H} = - J \sum_i \vec{\Sop}_{i} \cdot \vec{\Sop}_{i+1}$, and it is essentially related to the global rotational invariance of the system. It is named SU(2) symmetry since operators $\hat{S}^x, \hat{S}^y, \hat{S}^z$ satisfy the SU(2) commutation relations.

\chapter{Sixth lecture}
\section{The XXZ Hamiltonian with symmetries}
We consider again the $XXZ$ model for $N$ quantum spins
\begin{align}
\begin{split}
    \hat{H} &= - J \sum_{i=0}^{N-1} \left( \frac{1}{2} \left( \pauliplus_{i} \pauliminus_{i+1} + \pauliminus_{i} \pauliplus_{i+1} \right) + \frac{\Delta}{4} \pauliz_{i} \pauliz_{i+1} \right)\, \, .
\end{split}  
\end{align}
We want to exploit the $U(1)$ symmetry of the model, i.e.\ the fact that $\hat{H}$ is block diagonal, each block being labelled by the value of the total $z$ magnetization $\hat{S}^z$ (see Eq.\ref{eq:hamblockdiagonal}). The task is to design an Algorithm similar to \ref{alg:XXZ_Ham} to build $\hat{H}$ in a given magnetization sector. First, which is the dimension of these blocks/sectors? Well, consider our basis states $\ket{\spin_1, \spin_2, \ldots, \spin_N}$. If this state has $n_0$ spins being $\uparrow$, and $n_1$ spins being $\downarrow$,
the total magnetization will be 
\begin{equation}
   \hat{S}^z = \frac{1}{2} (n_0 - n_1). 
\end{equation}
Notice that $n_0$, $n_1$ are respectively the number of $0$ and $1$ in the bit representation of $\ket{\spin_1, \spin_2, \ldots, \spin_N}$. Also notice that $n_0 + n_1 = N$, since there $N$ spins in total, and therefore 
\begin{equation}
   \hat{S}^z = \frac{1}{2} (2 n_0 - N) = n_0 - \frac{N}{2} \, , 
\end{equation}
\begin{equation}
   n_0 = \hat{S}^z + \frac{N}{2} \, . 
\end{equation}
Now we want to count the number of states in this sector. The question is: how to arrange into different strings $n_0$ bits $0$ and $n_1 = N-n_0$ bits 1? The answer is given by the binomial coefficient:
\begin{equation}
    D=\binom{N}{n_0} = \frac{N!}{n_0! (N- n_0)!} = \frac{N!}{n_0! n_1!} \, .
\end{equation}
As a check, let us verify that when we sum these numbers over all blocks, i.e.\ over all values of $n_0=0,1,2, ... N$ we obtain the dimension of the total Hilbert space, which we know is $2^N$. In fact, we have
\begin{equation}
    \sum_{n_0=0}^{N} \binom{N}{n_0} = \sum_{n_0=0}^{N} \frac{N!}{n_0! (N-n_0)!} = \sum_{n_0=0}^{N} 1^{n_0}  1^{N-n_0} \frac{N!}{n_0! (N-n_0)!} = 2^N \, ,
\end{equation}
where we used Newton's binomial formula
\begin{equation}
    (x+y)^N = = \sum_{n=0}^{N} x^{n} y^{N-n} \frac{N!}{n! (N-n)!} \, .
\end{equation}
The dimension of a single block is significantly smaller than that of the entire Hamiltonian. For instance, in the case where $N=16$. $\hat{H}$ is a matrix with $2^N = 65,536$ columns and rows. However, the central block corresponding to $\hat{S}^z=0$ is much smaller, with a dimension of $D=\binom{N}{N/2}=12,870$. \\

Now we need to design an algorithm to collect all binary strings containing $n_0$ zeros. The algorithm iterates over all integers 
$x$ in the range $[0, 2^N-1]$, converts each integer to a binary string, and then checks the number of zeros. This can be done by summing the binary digits and verifying if the count of zeros equals $n_0$. The procedure is summarized in Algorithm \ref{alg:XXZ_states_magn}.

\begin{algorithm}[h!]
\caption{Function $\texttt{states\_generation}$ generating states with a given number of bits $0$}\label{alg:XXZ_states_magn}
\begin{flushleft}
\hspace*{\algorithmicindent} \textbf{Input}: the system size $N$, the number $n_0$ of bits that must be $0$
\end{flushleft}
\begin{algorithmic}[1]
\State Initialize an empty list: $\text{states}= \{ \, \}$
\For{($x=0$, $x=2^N-1$, $x++$)}
     \State Set $b$ to be the binary form of $x$ with $N$ digits: $b = \texttt{bit}(x,N)$
     \State $n=\texttt{sum\_array}(b)$
      \If{$n=n_0$} 
               \State append $x$ to $\text{states}$
      \EndIf 
\EndFor
\end{algorithmic}
\begin{flushleft}
\hspace*{\algorithmicindent} \textbf{Output}: the list $\text{states}$ containing all possible states with exactly $n_0$ bits equal to $0$, encoded as integers 
\end{flushleft}
\end{algorithm}

The output of Algorithm \ref{alg:XXZ_states_magn} is a sorted list of integers of length $\binom{N}{n_0}$. Now, we should have a function able to find an element (an integer) in this list. To do this we can exploit the fact that integers are in ascending order, therefore using the \textit{binary search algorithm} (see Ref. \cite{enwiki:1228829178}). 

Binary search (Algorithm \ref{alg:find_position}) compares the target value to the middle element of the array. If they are not equal, it eliminates the half where the target cannot be and continues searching in the remaining half. This process is repeated, each time comparing the target to the middle element of the current section, until the target value is found. Binary search is very fast, taking at most logarithmic time relative to the size of the array.

\begin{algorithm}[h!]
\caption{Function $\texttt{find\_position}$ to find the position of a given integer in a sorted array}\label{alg:find_position}
\begin{flushleft}
\hspace*{\algorithmicindent} \textbf{Input}: an array $\text{sorted\_list}$ containing integers sorted in ascending order, a number $\text{target}$ to find
\end{flushleft}
\begin{algorithmic}[1]
\State Initialize $i_{\text{min}} = 0$ and  $i_{\text{max}} = \texttt{length}(\text{sorted\_list}) - 1$
\While{$i_{\text{min}} \leq i_{\text{max}}$}
    \State Set $\text{mid} = \left\lfloor \frac{i_{\text{min}} + i_{\text{max}}}{2} \right\rfloor$
    \If{$\text{sorted\_list}[\text{mid}] = \text{target}$}
        \State \Return $\text{mid}$
    \ElsIf{$\text{sorted\_list}[\text{mid}] < \text{target}$}
        \State Set $i_{\text{min}} = \text{mid} + 1$
    \Else
        \State Set $i_{\text{max}} = \text{mid} - 1$
    \EndIf
\EndWhile
\State \Return $-1$ \Comment{Return -1 if the target is not found}
\end{algorithmic}
\begin{flushleft}
\hspace*{\algorithmicindent} \textbf{Output}: the position of $\text{target}$ in the array, or $-1$ if $\text{target}$ is not found
\end{flushleft}
\end{algorithm}

With the functions \texttt{states\_generation} and \texttt{find\_position} written, we can now develop an algorithm to define the XXZ Hamiltonian in a magnetization block labelled by the total magnetization $S^z$ (or equivalently by $n_0 = S^z + \frac{N}{2}$). This procedure is a slight variation of Algorithm \ref{alg:XXZ_Ham}, but outputs a matrix of dimension $D \times D$ instead of $2^N \times 2^N$.

\begin{algorithm}[H]
\caption{Generation of the XXZ Hamiltonian in a given magnetization block}\label{alg:XXZ_Ham_magn}
\begin{flushleft}
\hspace*{\algorithmicindent} \textbf{Input}: the system size $N$, the parameters $J, \Delta$ and the total magnetization $S^z$
\end{flushleft}
\begin{algorithmic}[1]
\State Set $n_0 = S^z + \frac{N}{2}$ and $D=\binom{N}{n_0}$
\State Initialize $H$ as a $D \times D$ matrix with all entries $0$ (or a sparse matrix)
\State Create the array of all possible states in the given magnetization block, each labelled by the corresponding integer: $\text{states} = \texttt{states\_generation}(N,n_0)$
\For{($x=0$, $x=D-1$, $x++$)}
     \State Extract the $x$-th state from the list: $w=\text{states}[x]$
     \State Set $b$ to be the binary form of $w$ with $N$ digits: $b = \texttt{bit}(w,N)$
     \For{($i=0$, $i=N-1$, $i++$)}
           \State Set $j = \mod(i+1, N)$
           \If{$b[i]=b[j]$} 
               \State Do $H[x,x] = H[x,x] - \frac{J \Delta}{4}$
            \Else
               \State Do $H[x,x] = H[x,x] + \frac{J \Delta}{4}$
               \State Flip digits $i,j$ in $b$ obtaining $b' = \texttt{flip}(b,i,j)$. 
               \State Set: $y = \texttt{num}(b',N)$
               \State Find position of $y$: $z=\texttt{find\_position}(y,\text{states})$ 
               \State Do $H[x,z] = H[x,z] - \frac{J}{2}$
\EndIf 
     \EndFor
\EndFor
\end{algorithmic}
\begin{flushleft}
\hspace*{\algorithmicindent} \textbf{Output}: the $XXZ$ Hamiltonian matrix $H$
\end{flushleft}
\end{algorithm}
\chapter{Seventh lecture}
\section{The graphical language of Tensors}
First, we introduce a graphical way of describing vectors, matrices, and more generally \textit{tensors}.

For our purpose, a vector $v$ can be considered as a simple list of numbers. The entries can be labelled with a single index $i$ which runs from $1$ to $n$, where $n$ is the length of the vector. Therefore, $v_i$ represents the $i$-th entry of the vector. We will represent vectors as shapes with one leg, which represents the index $i$, i.e.\ as:
\begin{equation}
v_{i} = \hspace{-0.5 mm}
\raisebox{0.35cm}{
   \begin{tikzpicture}[baseline=(current  bounding  box.center),scale=0.6]
   \definecolor{mycolor}{rgb}{0.82,0.82,1.}
    \draw[thick, black] (0,0) -- (0,1.2);
    \draw[thick, fill=mycolor] (0,0) circle (0.5);
    \node[scale=0.8] at (0,0) {\Large $v$};
    \node[scale=0.8] at (0.4,1.0) {\Large $i$};
\end{tikzpicture}
}
\end{equation}
Now, a matrix $M$ is a list of numbers labelled by two indices: an index $i$ for the row, and an index $j$ for the columns. The first runs from $1$ to $n$ (number of rows), the second from $1$ to $m$ (number of rows).  We will represent matrix $M$ as a shape with two legs, which represent the two indices $i$ and $j$, i.e.\ as:
\begin{equation}
M_{ij} = \hspace{-2 mm}
\raisebox{0.3cm}{
   \begin{tikzpicture}[baseline=(current  bounding  box.center),scale=0.6]
   \definecolor{mycolor}{rgb}{0.82,0.82,1.}
    \draw[thick, black] (0,0) -- (-0.8,1.4);
    \draw[thick, black] (0,0) -- (+0.8,1.4);
    \draw[thick, fill=mycolor] (0,0) circle (0.7);
    \node[scale=0.8] at (0,0) {\Large $M$};
    \node[scale=0.8] at (-0.95,1.0) {\Large $i$};
    \node[scale=0.8] at (+0.95,1.0) {\Large $j$};
\end{tikzpicture}
}
\end{equation}
A very important matrix is the identity matrix $\mathbb{1}$
which has matrix elements $\mathbb{1}_{ij} = \delta_{ij}$.
This will be represented as a line:
\begin{equation}
\delta_{ij} = \hspace{-0.5 mm}
\raisebox{0.05cm}{
   \begin{tikzpicture}[baseline=(current  bounding  box.center),scale=0.5]
   \definecolor{mycolor}{rgb}{0.82,0.82,1.}
    \draw[thick, black] (0,0) -- (0,1.2);
    \node[scale=0.6] at (0.4,1.0) {\Large $i$};
    \node[scale=0.6] at (0.4,0.0) {\Large $j$};
\end{tikzpicture}
}
\end{equation}
This graphical language can be generalized to tensors with arbitrary rank, i.e.\ arbitrary number of indices. For instance a rank-3 tensors will be represented as:
\begin{equation}
T_{i_1 i_2 i_3} = \hspace{-4 mm}
\raisebox{-0.1cm}{
   \begin{tikzpicture}[baseline=(current  bounding  box.center),scale=0.5]
   \definecolor{mycolor}{rgb}{0.82,0.82,1.}
    \draw[thick, black] (0,0) -- (1.5*0.866, 1.5*0.5);
    \draw[thick, black] (0,0) -- (-1.5*0.866, 1.5*0.5);
    \draw[thick, black] (0,0) -- (0, -1.5*1);  
    \node[scale=0.6] at (1.95*0.866, 1.95*0.5) {\Large $i_2$};
    \node[scale=0.6] at (0, -1.95*1) {\Large $i_3$};
    \node[scale=0.6] at (-1.95*0.866, 1.95*0.5) {\Large $i_1$};
    \draw[thick, fill=mycolor] (0,0) circle (0.6);  
    \node[scale=0.6] at (0,0) {\Large $T$};
\end{tikzpicture}
}
\end{equation}
Now, given two vectors $v$ and $w$ of the same size, we might want to represent their scalar product $v \cdot w = \sum_i v_i w_i$. This is done simply by joining the legs of $v$ and $w$, i.e.
\begin{equation}
v \cdot w = \sum_i v_i w_i = \hspace{-0.6 mm}
\raisebox{0.05cm}{
   \begin{tikzpicture}[baseline=(current  bounding  box.center),scale=0.5]
   \definecolor{mycolor}{rgb}{0.82,0.82,1.}
    \draw[thick, black] (0,0) -- (0,1.8);
    \draw[thick, fill=mycolor] (0,0) circle (0.5);
    \draw[thick, fill=mycolor] (0,1.8) circle (0.5);
    \node[scale=0.6] at (0,0) {\Large $v$};
    \node[scale=0.6] at (0,1.8) {\Large $w$};
    \node[scale=0.6] at (0.4,0.9) {\Large $i$};
\end{tikzpicture}
}
\end{equation}
Observe that the result of this \textit{contraction} is a new shape with no external legs, so it is a rank-0 tensor, i.e.\ a number (as expected).

Given two matrices $M$ and $N$ we will represent their matrix matrix multiplication as
\begin{equation}
(M N)_{ij} = \sum_k M_{ik} N_{kj} = \hspace{-0.7 mm}
\raisebox{0.3cm}{
   \begin{tikzpicture}[baseline=(current  bounding  box.center),scale=0.5]
   \definecolor{mycolor}{rgb}{0.82,0.82,1.}
    \draw[domain=0:3,samples=50,color=black,thick] plot (\x, -\x^2/1.5^2-1 +3*\x/1.5^2 +1);
    \draw[thick, black] (3,0) -- (+3.5,1.2);
    \draw[thick, black] (0,0) -- (-0.5,1.2);
    \draw[thick, fill=mycolor] (0,0) circle (0.5);
    \draw[thick, fill=mycolor] (3,0) circle (0.5);
    \node[scale=0.6] at (0,0) {\Large $M$};
    \node[scale=0.6] at (3,0) {\Large $N$};
    \node[scale=0.6] at (-0.7,1.0) {\Large $i$};
    \node[scale=0.6] at (1.5,1.35) {\Large $k$};
    \node[scale=0.6] at (3.7,1.0) {\Large $j$};
\end{tikzpicture}
}
\end{equation}
We can generalize this notation to the case in which two arbitrary tensors are contracted in some ways. For instance:
\begin{equation}
(T W)_{i_1 i_3 i_5} = \sum_{i_2, i_4} T_{i_1 i_2 i_4 i_5} W_{i_2 i_3 i_4} = \hspace{-0.7 mm}
\raisebox{0.cm}{
   \begin{tikzpicture}[baseline=(current  bounding  box.center),scale=0.5]
   \definecolor{mycolor}{rgb}{0.82,0.82,1.}
    \draw[domain=0:3,samples=50,color=black,thick] plot (\x, -\x^2/1.5^2-1 +3*\x/1.5^2 +1);
    \draw[domain=0:3,samples=50,color=black,thick] plot (\x, +\x^2/1.5^2 +1 -3*\x/1.5^2 -1);
    \draw[thick, black] (3,0) -- (3+1.5*1, 0);

    \draw[thick, black] (0,0) -- (-1.5*0.866, 1.5*0.5);
    \draw[thick, black] (0,0) -- (-1.5*0.866, -1.5*0.5);  
    \node[scale=0.6] at (-1.95*0.866, 1.95*0.5) {\Large $i_1$};
    \node[scale=0.6] at (1.5, 1.4) {\Large $i_2$};
    \node[scale=0.6] at (4, 0.7) {\Large $i_3$};
    \node[scale=0.6] at (1.5, -1.4) {\Large $i_4$};
    \node[scale=0.6] at (-1.95*0.866, -1.95*0.5) {\Large $i_5$}; 
    \draw[thick, fill=mycolor] (0,0) circle (0.6);
    \draw[thick, fill=mycolor] (3,0) circle (0.6);    
    \node[scale=0.6] at (0,0) {\Large $T$};
    \node[scale=0.6] at (3,0) {\Large $W$};
\end{tikzpicture}
}
\end{equation}
The tensor product of two vectors $T = v \otimes w$ can be represented simply by placing the two diagrams representing 
$v$ and $w$ side by side, as shown below 
\begin{equation}\label{eq:tensor_product}
\raisebox{0.25cm}{
   \begin{tikzpicture}[baseline=(current  bounding  box.center),scale=0.5]
   \definecolor{mycolor}{rgb}{0.82,0.82,1.}
       \draw[thick, black] (0.5,0) -- (0.5,1.8);
       \draw[thick, black] (1.5,0) -- (1.5,1.8);
       \node[scale=0.6] at (0.5,2.2) {\Large $i$};
       \node[scale=0.6] at (1.5,2.2) {\Large $j$};
       \draw[thick, fill=mycolor] (0,0) rectangle (2,1);
       \node[scale=0.6] at (1,0.5) {\Large $T$};

\end{tikzpicture}
}
\quad
=
\quad
\raisebox{0.25cm}{
   \begin{tikzpicture}[baseline=(current  bounding  box.center),scale=0.5]
   \definecolor{mycolor}{rgb}{0.82,0.82,1.}
    \draw[thick, black] (0,0) -- (0,1.2);
    \draw[thick, fill=mycolor] (0,0) circle (0.5);
    \node[scale=0.6] at (0,0) {\Large $v$};
    \node[scale=0.6] at (0.4,1.0) {\Large $i$};
\end{tikzpicture}
}
\quad 
\raisebox{0.25cm}{
   \begin{tikzpicture}[baseline=(current  bounding  box.center),scale=0.5]
   \definecolor{mycolor}{rgb}{0.82,0.82,1.}
    \draw[thick, black] (0,0) -- (0,1.2);
    \draw[thick, fill=mycolor] (0,0) circle (0.5);
    \node[scale=0.6] at (0,0) {\Large $w$};
    \node[scale=0.6] at (0.4,1.0) {\Large $j$};
\end{tikzpicture}
}
\end{equation}
In fact, the components of $T$ are just the product of the components of $v$ and $w$: $T_{ij} = v_i w_j$. These simple instruments are already enough to visualize some simple relations in linear algebra. For instance, we can easily proove that $\Tr[(A_1 \otimes B_1 ) (A_2 \otimes B_2 )] = \Tr[A_1 A_2] \Tr[B_1 B_2]$. In fact if we define the tensors $T_1 = A_1 \otimes B_1$ and $T_2 = A_2 \otimes B_2$, we have
\begin{equation}
\Tr[T_1 T_2] = 
\raisebox{0.10cm}{
   \begin{tikzpicture}[baseline=(current  bounding  box.center),scale=0.5]
   \definecolor{mycolor}{rgb}{0.82,0.82,1.}
       \draw[thick, black] (-0.5,3.3) -- (0.5,3.3);
       \draw[thick, black] (0.5,-0.5) -- (0.5,3.3);
       \draw[thick, black] (-0.5,-0.5) -- (0.5,-0.5);       
       \draw[thick, black] (1.5,-0.5) -- (1.5,3.3);
       \draw[thick, black] (1.5,3.3) -- (2.5,3.3);
       \draw[thick, black] (1.5,-0.5) -- (2.5,-0.5);
       \draw[thick, black] (-0.5,-0.5) -- (-0.5,3.3);
       \draw[thick, black] (2.5,-0.5) -- (2.5,3.3);
       \draw[thick, fill=mycolor] (0,0) rectangle (2,1);
       \node[scale=0.6] at (1,0.5) {\Large $T_1$};
       \draw[thick, fill=mycolor] (0,1.8) rectangle (2,2.8);
       \node[scale=0.6] at (1,2.3) {\Large $T_2$};

\end{tikzpicture}
} \, ,
\end{equation}
and, by using Eq. \ref{eq:tensor_product}, we get
\begin{equation}
\Tr[T_1 T_2] =
\raisebox{0.15cm}{
   \begin{tikzpicture}[baseline=(current  bounding  box.center),scale=0.6]
   \definecolor{mycolor}{rgb}{0.82,0.82,1.}
    \draw[thick, black] (0,-1.) -- (0,2.5);
    \draw[thick, black] (-1.,-1.) -- (-1.,2.5);
    \draw[thick, black] (-1.,2.5) -- (0.,2.5);
    \draw[thick, black] (-1.,-1.) -- (0.,-1.);
    \draw[thick, fill=mycolor] (0,0) circle (0.5);
    \node[scale=0.5] at (0,0) {\Large $A_1$};
    \draw[thick, fill=mycolor] (0,1.5) circle (0.5);
    \node[scale=0.5] at (0,1.5) {\Large $A_2$};  
\end{tikzpicture}
}
\raisebox{0.15cm}{
   \begin{tikzpicture}[baseline=(current  bounding  box.center),scale=0.6]
   \definecolor{mycolor}{rgb}{0.82,0.82,1.}
    \draw[thick, black] (0,-1.) -- (0,2.5);
    \draw[thick, black] (1,-1.) -- (1,2.5);
    \draw[thick, black] (1.,2.5) -- (0.,2.5);
    \draw[thick, black] (1.,-1.) -- (0.,-1.);
    \draw[thick, fill=mycolor] (0,0) circle (0.5);
    \node[scale=0.5] at (0,0) {\Large $B_1$};
    \draw[thick, fill=mycolor] (0,1.5) circle (0.5);
    \node[scale=0.5] at (0,1.5) {\Large $B_2$};    
\end{tikzpicture}
}
= \Tr[A_1 A_2] \Tr[B_1 B_2] \, .
\end{equation}

\section{Tensors in the quantum world}

The graphical language of tensors that we have introduced can be very useful in describing quantum many-body systems. For example, if we have a system of $N$ spins in a state $\ket{\psi}$, we can always decompose $\psi$ in the computational basis as follows:
\begin{equation}
    \ket{\psi} = \sum_{\spin_1, \spin_2, \ldots, \spin_N} \psi_{\spin_1, \spin_2, \ldots, \spin_N} \ket{\spin_1, \spin_2, \ldots, \spin_N} \, ,
\end{equation}
where $\sigma_i \in \{ 0, 1\}$ (or equivalently $\sigma_i \in \{\uparrow, \downarrow \}$), and $\ket{\spin_1, \spin_2, \ldots, \spin_N}$ represents a state of the computational basis. The set of complex numbers $\psi_{\spin_1, \spin_2, \ldots, \spin_N}$ can be thought as a rank-$N$ tensor, where the $\sigma_i$ are indices of the tensor. We can therefore represent $\psi$ as a shape with $N$ legs:
\begin{equation}
\psi_{\spin_1, \spin_2, \ldots, \spin_N} = 
\raisebox{0.45cm}{
   \begin{tikzpicture}[baseline=(current  bounding  box.center),scale=0.6]
   \definecolor{mycolor}{rgb}{0.82,0.82,1.}
   \pgfmathsetmacro{\ll}{0.8}
    \foreach \x in {0,...,6}{
        \draw[thick, black] (\x*\ll - 2.4 ,0) -- (\x*\ll - 2.4,1.9);
        \pgfmathsetmacro{\y}{int(\x + 1)}
        \node[scale=0.8] at (\ll*\x - 2.4, 2.25) {$\Large \ifnum\y<4
            \spin_{\y} 
        \else
            \ifnum\y>6
                \spin_{N}
            \else
               .
            \fi
        \fi$};
    }
   \draw[thick, fill=mycolor] (-2.85,-0.75) rectangle (2.85,0.75);
    \node[scale=0.8] at (0,0.) {\Large $\psi$};
\end{tikzpicture}
} \, .
\end{equation}
As we have seen, the complexity of this object for many-particle (spin) systems is formidable, and in general, even for modest values of $N$, this tensor cannot be processed or stored on a computer. However, in some cases, it is possible to find an approximate decomposition of the tensor $\psi$ that greatly reduces the costs of computations. 
\chapter{Eighth lecture}

\section{Tensor decomposition of the GHZ state}

Let us start by considering a specific case, a system of only $N=2$ spins in the following state:
\begin{equation}\label{eq:ghz2}
    \ket{\psi} = \frac{\ket{\uparrow \uparrow} + \ket{\downarrow \downarrow}}{\sqrt{2}}  = \frac{\ket{00} + \ket{11}}{\sqrt{2}} \, .
\end{equation}
The $2^N = 4$ entries of the tensor $\psi$ are:
\begin{equation}\label{eq:entries_psi}
    \psi_{00} = 2^{-1/2} \, , \quad  \psi_{01} = 0 \, , \quad \psi_{10} = 0  \, , \quad  \psi_{11} = 2^{-1/2} \, ,
\end{equation}
and the tensor can be represented as
\begin{equation}
\psi_{\spin_1, \spin_2} = 
\raisebox{0.6cm}{
   \begin{tikzpicture}[baseline=(current  bounding  box.center),scale=0.6]
   \definecolor{mycolor}{rgb}{0.82,0.82,1.}
   \pgfmathsetmacro{\ll}{0.8}
    \foreach \x in {0,...,1}{
        \draw[thick, black] (\x*\ll - 2.4 ,0) -- (\x*\ll - 2.4,1.9);
        \pgfmathsetmacro{\y}{int(\x + 1)}
        \node[scale=0.8] at (\ll*\x - 2.4, 2.25) {$\Large \ifnum\y<4
            \spin_{\y} 
        \else
            \ifnum\y>6
                \spin_{N}
            \else
               .
            \fi
        \fi$};
    }
   \draw[thick, fill=mycolor] (-2.85,-0.75) rectangle (-1.2,0.75);
\end{tikzpicture}
} \, .
\end{equation}
Now we consider the first spin and we associate to it two raw vectors of length $2$ defined as  
\begin{equation}
    A^{(1)}(\uparrow) = 2^{-1/4} \left( 1 , \, 0 \right) \qquad A^{(1)}(\downarrow) = 2^{-1/4} \left( 0 , \, 1 \right) \, .
\end{equation}
In the same way, we associate two column vectors of length $2$ to the second spin
\begin{equation}
A^{(2)}(\uparrow) = 
2^{-1/4} \left(
\begin{array}{c}
 1 \\
 0 \\
\end{array}
\right) \qquad A^{(2)}(\downarrow) = 
2^{-1/4} \left(
\begin{array}{c}
 0 \\
 1 \\
\end{array}
\right) 
\end{equation}
Vectors $A^{(1)}, A^{(2)}$ should be thought as \textit{spin dependent tensors}, in the sense that for each value of $\sigma_i$ we get a different vector. Therefore we can represent them as follows
\begin{equation}\label{eq:ghz_n2_tensor}
A^{(1)}_{\spin_1,a} = 
\raisebox{0.6cm}{
   \begin{tikzpicture}[baseline=(current  bounding  box.center),scale=0.6]
   \definecolor{mycolor}{rgb}{0.82,0.82,1.}
   \pgfmathsetmacro{\ll}{1.2}
   \draw[thick, black] (\ll - 2.4 ,0) -- (\ll - 2.4,1.6);
    \draw[line width=0.5mm, blue!99!black] (\ll-2.4,0) -- (\ll-2.4+1.5,0);
   \draw[thick, fill=mycolor] (\ll -2.4,0.) circle (0.75);
   \node[scale=0.8] at (\ll - 2.4, 0) {$\Large A^{(i)}$};
   \node[scale=0.8] at (\ll - 2.4, 1.85) {$\Large \spin_{1} $}; 
   \node[scale=0.8] at (\ll - 2.4 + 2., 0) {$\Large a $};
\end{tikzpicture}
}
\qquad \qquad 
\raisebox{0.6cm}{
   \begin{tikzpicture}[baseline=(current  bounding  box.center),scale=0.6]
   \definecolor{mycolor}{rgb}{0.82,0.82,1.}
   \pgfmathsetmacro{\ll}{1.2}
   \draw[thick, black] (\ll - 2.4 ,0) -- (\ll - 2.4,1.6);
    \draw[line width=0.5mm, blue!99!black] (\ll-2.4-1.5,0) -- (\ll-2.40,0);
   \draw[thick, fill=mycolor] (\ll -2.4,0.) circle (0.75);
   \node[scale=0.8] at (\ll - 2.4, 0) {$\Large A^{(i)}$};
   \node[scale=0.8] at (\ll - 2.4, 1.85) {$\Large \spin_{2} $};
   \node[scale=0.8] at (\ll - 2.4 - 2., 0) {$\Large a $};
\end{tikzpicture}
} = A^{(2)}_{a,\spin_2} \, ,
\end{equation}
where the index $a \in \{1,2\}$ labels the two entries of the vectors. Notice that $A^{(1)}_{\spin_1,a}$ is the same thing of $A^{(1)}_{a}(\spin_1)$, and $A^{(2)}_{a,\spin_2}$ is the same thing of $A^{(2)}_{a}(\spin_2)$. 
Now, since we defined raw vectors and columns vectors, let us compute their scalar products:
\begin{align}
    \begin{split}
        A^{(1)}(\uparrow) \cdot A^{(2)}(\uparrow) &= 2^{-1/2} \\
        A^{(1)}(\uparrow) \cdot A^{(2)}(\downarrow) &= 0\\
        A^{(1)}(\downarrow) \cdot A^{(2)}(\uparrow) &= 0\\
        A^{(1)}(\downarrow) \cdot A^{(2)}(\downarrow) &= 2^{-1/2} \\
    \end{split} \, .
\end{align}
The interesting thing is that these values coincide with the entries of the tensor $\psi$ (Eq.~\ref{eq:entries_psi}):
\begin{align}
    \begin{split}
        \psi_{00} = \braket{\uparrow \uparrow|\psi} &= A^{(1)}(\uparrow) \cdot A^{(2)}(\uparrow) \\
        \psi_{01} = \braket{\uparrow \downarrow|\psi} &= A^{(1)}(\uparrow) \cdot A^{(2)}(\downarrow) \\
        \psi_{10} = \braket{\downarrow \uparrow|\psi} &= A^{(1)}(\downarrow) \cdot A^{(2)}(\uparrow) \\
        \psi_{11} = \braket{\downarrow \downarrow|\psi} &= A^{(1)}(\downarrow) \cdot A^{(2)}(\downarrow) \\
    \end{split}
\end{align}
We can therefore draw the conclusion that:
\begin{align}
    \begin{split}
        \psi_{\sigma_1 \sigma_2} = A^{(1)}(\sigma_1) \cdot A^{(2)}(\sigma_2) 
    \end{split}
\end{align}
or also, by rewriting explicitly the sum inside the scalar product, 
\begin{align}
    \begin{split}
        \psi_{\sigma_1 \sigma_2} = \sum_{a=1}^2 A^{(1)}_{\sigma_1 a} A^{(2)}_{a \sigma_2} \, .
    \end{split}
\end{align}
In the graphical language this can be represented with the following identity
\begin{equation}\label{eq:ghz_n2}
\raisebox{0.6cm}{
   \begin{tikzpicture}[baseline=(current  bounding  box.center),scale=0.6]
   \definecolor{mycolor}{rgb}{0.82,0.82,1.}
   \pgfmathsetmacro{\ll}{0.8}
    \foreach \x in {0,...,1}{
        \draw[thick, black] (\x*\ll - 2.4 ,0) -- (\x*\ll - 2.4,1.9);
        \pgfmathsetmacro{\y}{int(\x + 1)}
        \node[scale=0.8] at (\ll*\x - 2.4, 2.25) {$\Large \ifnum\y<4
            \spin_{\y} 
        \else
            \ifnum\y>6
                \spin_{N}
            \else
               .
            \fi
        \fi$};
    }
   \draw[thick, fill=mycolor] (-2.85,-0.75) rectangle (-1.2,0.75);
\end{tikzpicture}
} = 
\raisebox{0.6cm}{
   \begin{tikzpicture}[baseline=(current  bounding  box.center),scale=0.6]
   \definecolor{mycolor}{rgb}{0.82,0.82,1.}
   \pgfmathsetmacro{\ll}{1.5}
    \draw[line width=0.5mm, blue!99!black] (-2*\ll,0) -- (1*\ll-2*\ll,0);
    \foreach \x in {0,...,1}{
        \draw[thick, black] (\x*\ll - 2*\ll ,0) -- (\x*\ll - 2*\ll,1.5);
        \pgfmathsetmacro{\y}{int(\x + 1)}
        \node[scale=0.8] at (\ll*\x - 2*\ll, 1.85) {$\Large \ifnum\y<4
            \spin_{\y} 
        \else
            \ifnum\y>6
                \spin_{N}
            \else
               .
            \fi
        \fi$};
       \draw[thick, fill=mycolor] (\ll*\x - 2*\ll,0.) circle (0.45);
    }
\end{tikzpicture}
} \, ,
\end{equation}
where the blue legs are joined to represent the sum over the index $a$. Note that this index, which can take two values $a \in \{1, 2\}$, is somewhat unphysical, as it does not represent a spin. Instead, it is an auxiliary variable (or index) that we have introduced to decompose the original tensor $\psi$ into two parts. \\

At this point, we can try to generalize these considerations to the case of $N$ spins. To this purpose, let us consider the following state:
\begin{equation}
    \ket{\psi} = \frac{\ket{\uparrow \uparrow ... \uparrow} + \ket{\downarrow \downarrow  ... \downarrow}}{\sqrt{2}}  = \frac{\ket{00 ... 0} + \ket{11 ... 1}}{\sqrt{2}} 
\end{equation}
which is usually named Greenberger–Horne–Zeilinger (GHZ) state and reduces to the state Eq.~\ref{eq:ghz2} for $N=2$.
Our goal is to find a \textit{tensor decomposition} of the tensor $\psi$ similar to the one depicted in Eq.~\ref{eq:ghz_n2}. A natural extension of Eq.~\ref{eq:ghz_n2} to the case with $N$ particles is 
\begin{equation}\label{eq:mps}
\raisebox{0.6cm}{
   \begin{tikzpicture}[baseline=(current  bounding  box.center),scale=0.6]
   \definecolor{mycolor}{rgb}{0.82,0.82,1.}
   \pgfmathsetmacro{\ll}{0.8}
    \foreach \x in {0,...,6}{
        \draw[thick, black] (\x*\ll - 2.4 ,0) -- (\x*\ll - 2.4,1.9);
        \pgfmathsetmacro{\y}{int(\x + 1)}
        \node[scale=0.8] at (\ll*\x - 2.4, 2.25) {$\Large \ifnum\y<4
            \spin_{\y} 
        \else
            \ifnum\y>6
                \spin_{N}
            \else
               .
            \fi
        \fi$};
    }
   \draw[thick, fill=mycolor] (-2.85,-0.75) rectangle (2.85,0.75);
    \node[scale=0.6] at (0,0.) {\Large $\psi$};
\end{tikzpicture}
} = 
\raisebox{0.6cm}{
   \begin{tikzpicture}[baseline=(current  bounding  box.center),scale=0.6]
   \definecolor{mycolor}{rgb}{0.82,0.82,1.}
   \pgfmathsetmacro{\ll}{1.5}
    \draw[line width=0.5mm, blue!99!black] (-2*\ll,0) -- (6*\ll-2*\ll,0);
    \foreach \x in {0,...,6}{
        \draw[thick, black] (\x*\ll - 2*\ll ,0) -- (\x*\ll - 2*\ll,1.5);
        \pgfmathsetmacro{\y}{int(\x + 1)}
        \node[scale=0.8] at (\ll*\x - 2*\ll, 1.85) {$\Large \ifnum\y<4
            \spin_{\y} 
        \else
            \ifnum\y>6
                \spin_{N}
            \else
               .
            \fi
        \fi$};
       \draw[thick, fill=mycolor] (\ll*\x -2*\ll,0.) circle (0.45);
    }
\end{tikzpicture}
}
\end{equation}
Notice that at the boundaries (sites $i=1$ and $i=N$) we have tensors of the same shape of the one represented in Eq.~\ref{eq:ghz_n2_tensor}, i.e.\ rank 2 tensors, while in the bulk we have rank 3 tensors as the following:
\begin{equation}
A^{(i)}_{a_{i-1},\spin_i,a_{i}} = 
\raisebox{0.6cm}{
   \begin{tikzpicture}[baseline=(current  bounding  box.center),scale=0.6]
   \definecolor{mycolor}{rgb}{0.82,0.82,1.}
   \pgfmathsetmacro{\ll}{1.2}
   \draw[thick, black] (\ll - 2.4 ,0) -- (\ll - 2.4,1.6);
    \draw[line width=0.5mm, blue!99!black] (\ll-2.4-1.5,0) -- (\ll-2.4+1.5,0);
   \draw[thick, fill=mycolor] (\ll -2.4,0.) circle (0.75);
   \node[scale=0.8] at (\ll - 2.4, 0) {$\Large A^{(i)}$};
   \node[scale=0.8] at (\ll - 2.4, 1.85) {$\Large \spin_{i} $};
   \node[scale=0.8] at (\ll - 2.4 - 2.2, 0) {$\Large a_{i-1} $};
   \node[scale=0.8] at (\ll - 2.4 + 2., 0) {$\Large a_{i} $};
\end{tikzpicture}
}
\end{equation}

Analogous to the previous case, we define the tensors $A$ as follows:
\begin{equation}
    A^{(1)}(\uparrow) = 2^{-\frac{1}{2N}} \left( 1 , \, 0 \right) \qquad A^{(1)}(\downarrow) = 2^{-\frac{1}{2N}} \left( 0 , \, 1 \right) \qquad \text{          for site $1$} \, ,
\end{equation}
\begin{equation}\label{eq:ghz_n_bulk}
A^{(i)}(\uparrow) = 
2^{-\frac{1}{2N}} \left(
\begin{array}{cc}
 1 & 0 \\
 0 & 0 \\
\end{array}
\right) \quad A^{(i)}(\downarrow) = 
2^{-\frac{1}{2N}} \left(
\begin{array}{cc}
 0 & 0 \\
 0 & 1 \\
\end{array}
\right)  \quad \text{ for sites $1<i<N$} \, ,
\end{equation}
\begin{equation}
A^{(N)}(\uparrow) = 
2^{-\frac{1}{2N}} \left(
\begin{array}{c}
 1 \\
 0 \\
\end{array}
\right) \qquad A^{(N)}(\downarrow) = 
2^{-\frac{1}{2N}} \left(
\begin{array}{c}
 0 \\
 1 \\
\end{array}
\right)  \qquad \text{          for site $N$} \, ,
\end{equation}
The wave function tensor is now 
\begin{align}
    \begin{split}
        \psi_{\sigma_1 \sigma_2 ... \sigma_N} = A^{(1)}(\sigma_1) \cdot A^{(2)}(\sigma_2) \cdot ... \cdot A^{(N)}(\sigma_N) \, ,
    \end{split}
\end{align}
and one can easily verify that indeed this gives the GHZ state. If for instance we consider the case in which all spins are up (i.e.\ $\sigma_1 = \sigma_2 = ... = \sigma_N = \uparrow$), we will have

\begin{align}
    \begin{split}
        A^{(1)}(\uparrow) \cdot A^{(2)}(\uparrow) \cdot ... \cdot A^{(N-1)} \cdot A^{(N)}(\uparrow) &= 2^{-\frac{1}{2}} \left( 1 , \, 0 \right) \left( \begin{array}{cc}
 1 & 0 \\
 0 & 0 \\
\end{array} \right) ... \left( \begin{array}{cc}
 1 & 0 \\
 0 & 0 \\
\end{array} \right)
\left(
\begin{array}{c}
 1 \\
 0 \\
\end{array}
\right) = \\
&=2^{-\frac{1}{2}} \left( 1 , \, 0 \right) 
\left(
\begin{array}{c}
 1 \\
 0 \\
\end{array}
\right) = 2^{-\frac{1}{2}}
\end{split}
\end{align}
that is indeed equal to the wave function amplitude for this spin configuration: $\braket{\uparrow \uparrow ... \uparrow|\psi} = 1/\sqrt{2}$. \\ 

\section{Matrix Product States (MPS)}
After these introductory examples, we now arrive at an important concept: a quantum many-body state $\ket{\psi}$ is named a \textit{Matrix Product State (MPS)} if its wave function tensor can be decomposed into a product of rank-3 tensors (rank-2 tensor at the boundaries) arranged in a one-dimensional ``train'', as illustrated in Eq.~\ref{eq:mps}. The dimension of the indices labeled by $a_i$ (that is, the blue lines in Eq.~\ref{eq:mps}) is called the \textit{ bond dimension} of the MPS. For example, as shown in the previous section, the GHZ state on $N$ spins is an MPS. The bond dimension of this MPS is $2$, as the matrices $A^{(i)}(\uparrow)$ and $A^{(i)}(\downarrow)$ in Eq.~\ref{eq:ghz_n_bulk} are $2 \times 2$ matrices. \\

Importantly, it can be shown that \textit{any quantum state can be represented as an MPS, provided the bond dimension is large enough}. Specifically, for generic states of $N$ spin-1/2 particles, the bond dimension scales as $\sim 2^{N/2}$, i.e.\ is exponentially large in the system size $N$. 
This result may seem somewhat futile: indeed, our original goal was to find a way to reduce the exponentially large complexity of the tensor $\ket{\psi}$ into an efficient representation. However, if the MPS itself has an exponentially large bond dimension, it cannot constitute an efficient representation of the state. \\

However, although generic states 
$\ket{\psi}$ are MPS with exponentially large bond dimensions, \textit{there are many physically interesting states that can be represented as MPS with small bond dimensions}. For instance, it turns out that \textit{ground states of typical Hamiltonians in one dimensional spin chains} are MPS with finite bond dimension $\chi$, that remains independent of the system size $N$~\cite{Hastings_2007}. To be more specific, the statement holds for Hamiltonians $\hat{H}$ with a finite energy gap between the ground state and the first excited state, which exclude for instance critical systems (see Chapter \ref{ch:cpt} for details). This result is crucial and directly relates to the intrinsic quantum correlations of the state, specifically its \textit{entanglement}. Notably, for such ground states, the \textit{entanglement entropy} between a subsystem and its complement does not increase with the subsystem's size—a phenomenon known as the ``\textit{area law}''~\cite{Hastings_2007}.

\begin{figure}[h!]
\centering
\includegraphics[width=0.5\linewidth]{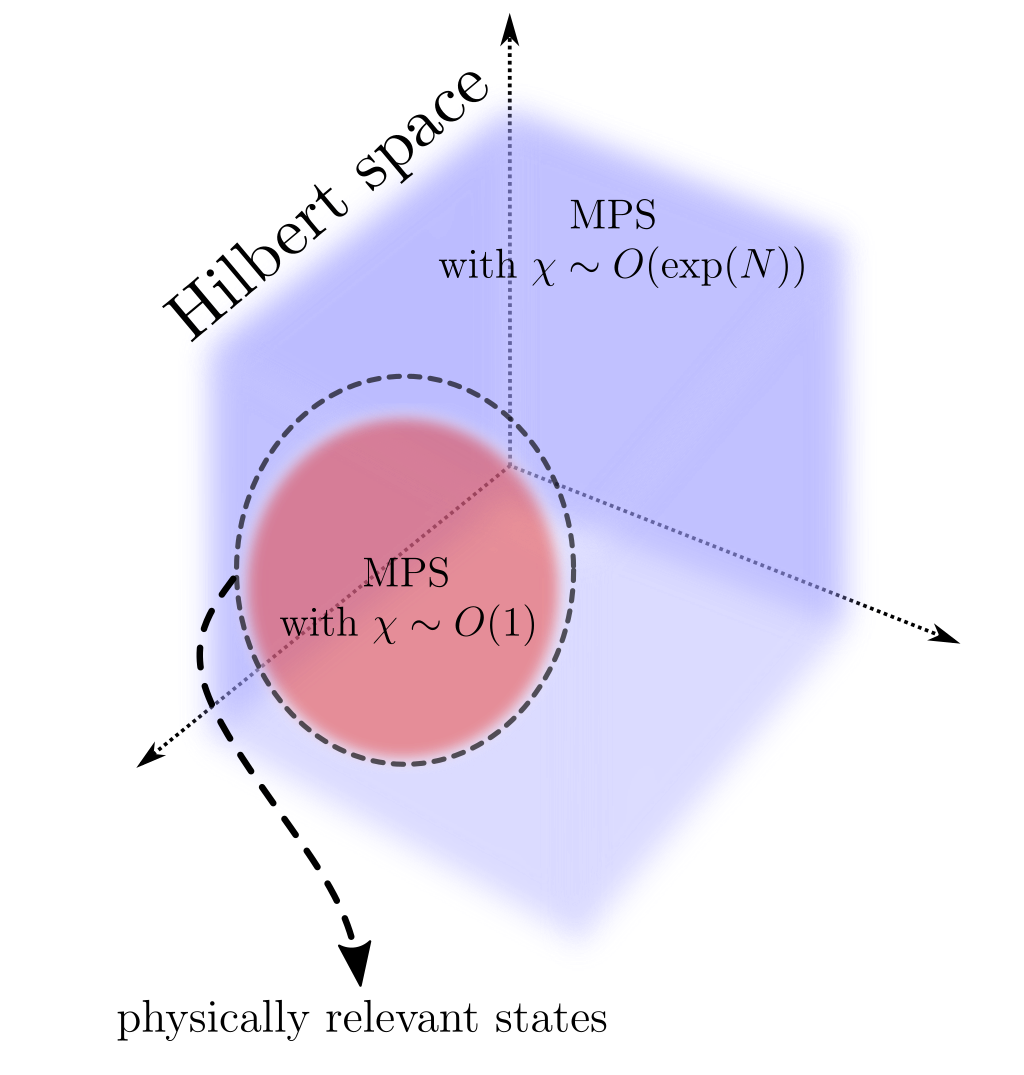}
\caption{Illustrative cartoon of the Hilbert space of a many-body system with $N$ spins. While generic states can be expressed in the form of a Matrix Product State (MPS) only at the price of an exponentially large bond dimension $\chi$, many physically relevant states live in a ``corner'' of the Hilbert space, and can be expressed as MPS with fixed bond dimension $\chi \sim O(1)$.} 
\label{fig:hs}
\end{figure}

\newpage
\section{Matrix Product Operators (MPO)}

The tensor decomposition we considered for states to arrive at the MPS representation can also be extended to the realm of operators. Let us therefore consider a generic operator $\hat{O}$ and represent it in the spin basis as follows
\begin{equation}
    \hat{O} = \sum_{\substack{\spin_1', \spin_2', \ldots, \spin_N' \\ \spin_1, \spin_2, \ldots, \spin_N}} O_{\spin_1', \spin_2', \ldots, \spin_N'; \spin_1, \spin_2, \ldots, \spin_N} \ket{\spin_1', \spin_2', \ldots, \spin_N'} \bra{\spin_1, \spin_2, \ldots, \spin_N} \, ,
\end{equation}
where $O_{\spin_1', \spin_2', \ldots, \spin_N'; \spin_1, \spin_2, \ldots, \spin_N}$ are essentially the entries of the matrix representing $\hat{O}$ in the basis of states $\ket{\spin_1, \spin_2, \ldots, \spin_N}$. As a tensor $O$ looks like a shape with $2N$ legs, as follows:
\begin{equation}\label{eq:mpo}
\raisebox{0.6cm}{
   \begin{tikzpicture}[baseline=(current  bounding  box.center),scale=0.6]
   \definecolor{mycolor}{rgb}{0.82,0.82,1.}
   \pgfmathsetmacro{\ll}{0.8}
    \foreach \x in {0,...,6}{
        \draw[thick, black] (\x*\ll - 2.4 ,-1.9) -- (\x*\ll - 2.4,1.9);
        \pgfmathsetmacro{\y}{int(\x + 1)}
        \node[scale=0.8] at (\ll*\x - 2.4, 2.25) {$\Large \ifnum\y<4
            \spin_{\y} 
        \else
            \ifnum\y>6
                \spin_{N}'
            \else
               .
            \fi
        \fi$};
        \node[scale=0.8] at (\ll*\x - 2.4, -2.25) {$\Large \ifnum\y<4
            \spin_{\y} 
        \else
            \ifnum\y>6
                \spin_{N}
            \else
               .
            \fi
        \fi$};
    }
   \draw[thick, fill=mycolor] (-2.85,-0.75) rectangle (2.85,0.75);
    \node[scale=0.8] at (0,0.) {\Large $O$};
\end{tikzpicture}
}
\end{equation}
A decomposition analogous to that of the tensor $\psi$ applies. Therefore, we can always decompose the tensor $O$ as follows:
\begin{equation}\label{eq:mpo}
\raisebox{0.3cm}{
   \begin{tikzpicture}[baseline=(current  bounding  box.center),scale=0.6]
   \definecolor{mycolor}{rgb}{0.82,0.82,1.}
   \pgfmathsetmacro{\ll}{0.8}
    \foreach \x in {0,...,6}{
        \draw[thick, black] (\x*\ll - 2.4 ,-1.9) -- (\x*\ll - 2.4,1.9);
        \pgfmathsetmacro{\y}{int(\x + 1)}
        \node[scale=0.8] at (\ll*\x - 2.4, 2.25) {$\Large \ifnum\y<4
            \spin_{\y} 
        \else
            \ifnum\y>6
                \spin_{N}'
            \else
               .
            \fi
        \fi$};
        \node[scale=0.8] at (\ll*\x - 2.4, -2.25) {$\Large \ifnum\y<4
            \spin_{\y} 
        \else
            \ifnum\y>6
                \spin_{N}
            \else
               .
            \fi
        \fi$};
    }
   \draw[thick, fill=mycolor] (-2.85,-0.75) rectangle (2.85,0.75);
    \node[scale=0.8] at (0,0.) {\Large $O$};
\end{tikzpicture}
} 
= 
\raisebox{0.3cm}{
   \begin{tikzpicture}[baseline=(current  bounding  box.center),scale=0.6]
   \definecolor{mycolor}{rgb}{0.82,0.82,1.}
   \pgfmathsetmacro{\ll}{1.5}
    \draw[line width=0.5mm, blue!99!black] (-2*\ll,0) -- (6*\ll-2*\ll,0);
    \foreach \x in {0,...,6}{
        \draw[thick, black] (\x*\ll - 2*\ll ,-1.5) -- (\x*\ll - 2*\ll,1.5);
        \pgfmathsetmacro{\y}{int(\x + 1)}
        \node[scale=0.8] at (\ll*\x - 2*\ll, 1.85) {$\Large \ifnum\y<4
            \spin_{\y}'
        \else
            \ifnum\y>6
                \spin_{N}
            \else
               .
            \fi
        \fi$};
        \node[scale=0.8] at (\ll*\x - 2*\ll, -1.85) {$\Large \ifnum\y<4
            \spin_{\y}
        \else
            \ifnum\y>6
                \spin_{N}
            \else
               .
            \fi
        \fi$};
       \draw[thick, fill=mycolor] (\ll*\x -2*\ll-0.45,-0.45) rectangle (\ll*\x -2*\ll+0.45,0.45);
    }
\end{tikzpicture}
} \, ,
\end{equation}
where the rectangles are rank-4 tensors (rank-3 tensors at the boundaries), as illustrated here
\begin{equation}
W^{(i)}_{a_{i-1},\spin_i',\spin_i,a_{i}} = 
\raisebox{0.25cm}{
   \begin{tikzpicture}[baseline=(current  bounding  box.center),scale=0.6]
   \definecolor{mycolor}{rgb}{0.82,0.82,1.}
   \pgfmathsetmacro{\ll}{1.2}
   \draw[thick, black] (\ll - 2.4 ,-1.6) -- (\ll - 2.4,1.6);
    \draw[line width=0.5mm, blue!99!black] (\ll-2.4-1.5,0) -- (\ll-2.4+1.5,0);
   \draw[thick, fill=mycolor] (\ll -2.4-0.75,-0.75) rectangle (\ll -2.4+0.75,0.75);
   \node[scale=0.8] at (\ll - 2.4, 0) {$\Large W^{(i)}$};
   \node[scale=0.8] at (\ll - 2.4, 1.85) {$\Large \spin_{i}' $};
   \node[scale=0.8] at (\ll - 2.4, -1.85) {$\Large \spin_{i} $};
   \node[scale=0.8] at (\ll - 2.4 - 2.2, 0) {$\Large a_{i-1} $};
   \node[scale=0.8] at (\ll - 2.4 + 2., 0) {$\Large a_{i} $};
\end{tikzpicture}
} \, .
\end{equation}
Notice that, as before, blue lines represent auxiliary indices $a_i \in \{1, 2, \dots, \chi\}$, where $\chi$ is the bond dimension, while black lines correspond to physical indices. The object on the right-hand side of Eq.~\ref{eq:mpo} is referred to as a Matrix Product Operator (MPO). The tensors $W^{(i)}$ can also be interpreted as matrices of operators, since fixing the indices $a_{i-1}$ and $a_{i}$ leaves a matrix with indices $\sigma_i'$ and $\sigma_i$, which acts on the physical spin $i$.

As with states, the most generic operator can also be represented as an MPO, but only at the cost of a large bond dimension $\chi$. However, many physically relevant observables are MPOs with small bond dimensions. For example, the otal magnetization operator
\begin{equation}
    \hat{M}_z = \sum_{i=1}^N  \pauliz_i
\end{equation}
is an MPO of bond dimension $2$. In fact, it is possible to verify that the matrices of operators 
\begin{equation}
W^{(i)} = 
\left(
\begin{array}{cc}
 \hat{\mathbb{1}}_i & 0 \\
 \pauliz_i &  \hat{\mathbb{1}}_i \\
\end{array}
\right)
\end{equation}
for $1<i<N$, together with the vectors 
\begin{equation}
W^{(1)} = 
\left(
\begin{array}{cc}
 \pauliz_1 &  \hat{\mathbb{1}}_1 \\
\end{array}
\right)
\qquad 
W^{(N)} = 
\left(
\begin{array}{cc}
 \hat{\mathbb{1}}_N \\
 \pauliz_N \\
\end{array}
\right)
\end{equation}
at the two boundaries do the job, namely $\hat{M}_z = W^{(1)} W^{(2)} ... W^{(N)}$.

MPS and MPO representations can be highly effective in numerical practice through the use of \textit{tensor network algorithms}~\cite{Collura2024, SCHOLLWOCK201196}. A simple example of how MPS and MPO can be combined is the calculation of the matrix times vector product $\hat{O} \ket{\psi}$, where $\hat{O}$ in an operator expressed as an MPO, and $\ket{\psi}$ is a state expressed as an MPS. In graphical notation, this appears as follows:
\begin{equation}\label{eq:mpo_to_mps}
 \hat{O} | \psi \rangle =
\raisebox{0.15cm}{
   \begin{tikzpicture}[baseline=(current  bounding  box.center),scale=0.4]
   \definecolor{mycolor}{rgb}{0.82,0.82,1.}
   \pgfmathsetmacro{\ll}{1.5}
    \draw[line width=0.5mm, blue!99!black] (-2*\ll,0) -- (6*\ll-2*\ll,0);
    \draw[line width=0.5mm, blue!99!black] (-2*\ll,-1.5) -- (6*\ll-2*\ll,-1.5);
    \foreach \x in {0,...,6}{
        \draw[thick, black] (\x*\ll - 2*\ll ,-1.5) -- (\x*\ll - 2*\ll,1.5);
        \pgfmathsetmacro{\y}{int(\x + 1)}
       \draw[thick, fill=mycolor] (\ll*\x -2*\ll-0.45,-0.45) rectangle (\ll*\x -2*\ll+0.45,0.45);
       \draw[thick, fill=mycolor] (\ll*\x -2*\ll,-1.5) circle (0.45);
    }
\end{tikzpicture}
}
\end{equation}
Similarly, the expectation value of an observable $\hat{O}$ on a state $\ket{\psi}$ appears as follows~\cite{SCHOLLWOCK201196}:

\begin{equation}\label{eq:mps_mpo_mps}
\langle \psi | \hat{O} | \psi \rangle =
\raisebox{0.15cm}{
   \begin{tikzpicture}[baseline=(current  bounding  box.center),scale=0.4]
   \definecolor{mycolor}{rgb}{0.82,0.82,1.}
   \pgfmathsetmacro{\ll}{1.5}
    \draw[line width=0.5mm, blue!99!black] (-2*\ll,0) -- (6*\ll-2*\ll,0);
    \draw[line width=0.5mm, blue!99!black] (-2*\ll,-1.5) -- (6*\ll-2*\ll,-1.5);
    \draw[line width=0.5mm, blue!99!black] (-2*\ll,+1.5) -- (6*\ll-2*\ll,+1.5);
    \foreach \x in {0,...,6}{
        \draw[thick, black] (\x*\ll - 2*\ll ,-1.5) -- (\x*\ll - 2*\ll,1.5);
        \pgfmathsetmacro{\y}{int(\x + 1)}
       \draw[thick, fill=mycolor] (\ll*\x -2*\ll-0.45,-0.45) rectangle (\ll*\x -2*\ll+0.45,0.45);
       \draw[thick, fill=mycolor] (\ll*\x -2*\ll,1.5) circle (0.45);
       \draw[thick, fill=mycolor] (\ll*\x -2*\ll,-1.5) circle (0.45);
    }
\end{tikzpicture}
}
\end{equation}
Both tensor network contractions of Eq.~\ref{eq:mpo_to_mps} and Eq.~\ref{eq:mps_mpo_mps} can be evaluated efficiently using suitable  Algorithms that performs the contractions in a specific, optimized way~\cite{SCHOLLWOCK201196}. 
For instance, the expectation value Eq.~\ref{eq:mps_mpo_mps} can be computed at a computational cost $O(N(\chi_{\text{MPS}}^3 \chi_{\text{MPO}} + \chi_{\text{MPS}}^2 \chi_{\text{MPO}}^2))$, where $\chi_{\text{MPS}}, \chi_{\text{MPO}}$ are respectively the bond dimensions of the MPS and of the MPO. Note that the cost is linear in $N$, whereas using the state and operator representations as a vector of length $2^N$ and a matrix of size $2^N \times 2^N$, respectively, would result in a cost of $O(\exp(N))$.
\chapter{Python Codes}\label{ch:code}
In this chapter, we will present the algorithms discussed in the main text as ready-to-use Python code. The codes will be organized in the same order as they appear in the previous chapters for easy reference and navigation.

\section{Codes for Chapter 4}

The following three Python functions implement the \texttt{bit} (\ref{eq:bin}), \texttt{num} (\ref{eq:num}), and \texttt{flip} (\ref{eq:flip}) functions, respectively. 

\hspace{1cm}

\begin{lstlisting}[style=pythonstyle,
caption={Python function to convert integer to a binary NumPy array}, captionpos=b, captionpos=b]
def bit(x: int, N: int) -> np.ndarray:
    """
    Converts an integer x into a binary NumPy array of length N.

    Arguments:
        x (int): The input integer.
        N (int): The desired length of the output array.

    Raises:
        ValueError: If the binary representation of x is longer than N.

    Returns:
        np.ndarray: A NumPy array of length N filled with 0s and 1s.
    """
    
    # Convert the integer x to its binary representation and remove the '0b' prefix
    binary_str = bin(x)[2:]
    
    # Check if the length of binary_str is greater than N
    if len(binary_str) > N:
        raise ValueError(f"The binary representation of {x} is longer than {N} bits.")
    
    # Pad the binary string with zeros at the beginning to ensure it is of length N
    binary_str = binary_str.zfill(N)
    
    # Convert the string to an array of integers
    binary_array = np.array([int(bit) for bit in binary_str])
    
    return binary_array
\end{lstlisting}

\begin{lstlisting}[style=pythonstyle, captionpos=b, caption={Python function to convert a binary NumPy array to an integer}]
def num(binary_array: np.ndarray) -> int:
    """
    Converts a NumPy array of binary digits to its corresponding integer.

    Arguments:
        binary_array (np.ndarray): A NumPy array of binary digits (0s and 1s).

    Returns:
        int: The integer corresponding to the binary representation.
    """
    
    # Convert the NumPy array to a list of strings and join them to form the binary string
    binary_str = ''.join(binary_array.astype(str))
    
    # Convert the binary string to an integer
    integer_value = int(binary_str, 2)
    
    return integer_value
\end{lstlisting}

\begin{lstlisting}[style=pythonstyle, captionpos=b, caption={Python function to flip elements in an array}]
def flip(arr: np.ndarray, i: int, j: int) -> np.ndarray:
    """
    Flips the elements at two specified positions in an array.

    Arguments:
        arr (list): The input list/array.
        i (int): The index of the first position to flip.
        j (int): The index of the second position to flip.

    Returns:
        list: The array with elements at pos1 and pos2 flipped.
    """
    
    # Ensure positions are within bounds
    if i < 0 or i >= len(arr) or j < 0 or j >= len(arr):
        raise IndexError("Positions are out of bounds.")
        
    arr_new = arr.copy()
    
    # Swap elements at pos1 and pos2
    arr_new[i], arr_new[j] = arr_new[j], arr_new[i]
    
    return arr_new
\end{lstlisting}

\hspace{2cm}

The following function corresponds to Algorithm \ref{alg:XXZ_Ham} to construct a XXZ hamiltonian

\begin{lstlisting}[style=pythonstyle, captionpos=b, caption={XXZ Hamiltonian Function in Python}]
def XXZ_Ham(N: int, J: float, delta: float) -> np.ndarray:
    """
    Builds the XXZ hamiltonian matrix in Z basis with PBC

    Arguments:
        N :  system size
        J : interaction strength
        delta : anisotropy strength 

    Returns:
        Ham: Full hamiltonian matrix of XXZ model
    """
    
    Ham = np.zeros((2**N, 2**N))  # initialize a 2^(N) by 2^(N) matrix filled with zeros
    
    for x in range(2**N):  # loop over Hilbert space 
        b = bit(x, N)  # convert x into its binary representation
                    
        for i in range(N):  # loop over sites
            j = np.mod(i+1, N)  # j is the right nearest neighbor of i in PBC
                        
            if b[i] == b[j]:
                Ham[x, x] += -J * delta / 4  # same config in i and j gives positive sign for ZZ term
            else:
                Ham[x, x] += J * delta / 4   # different config in i and j gives negative sign for ZZ term
                b_new = flip(b, i, j)         # apply spin flip term
                y = num(b_new)                # convert binary to integer
                Ham[x, y] += -J / 2           # fill the corresponding location
        
    return Ham
\end{lstlisting}

\hspace{1cm}

The following four functions correspond to Algorithms \ref{alg:X_i}, \ref{alg:Y_i}, \ref{alg:Z_i}, and \ref{alg:XX_i} respectively. These functions are used to generate a generic spin one-half Hamiltonians,

\hspace{1cm}

\begin{lstlisting}[style=pythonstyle, captionpos=b, caption={Function \texttt{X} that applies the Pauli-X operator}]
def X(x: int, i: int, N: int):
    """
    Applies the Pauli-X operator to the i-th site of the given state.
    
    Arguments:
        x (int): The input state as integer.
        i (int): The site index where the Pauli-X operator is applied (0-based index).
        N (int): System size
        
    Returns:
        tuple: The coefficient and the resultant state.
    """
    
    # Convert integer to binary array
    state = bit(x, N)
    
    # Apply the Pauli-X operator by flipping the i-th bit
    state[i] = 1 - state[i]
    
    # Determine the coefficient
    coef = 1.0

    # Convert state to integer
    y = num(state)   
    
    return coef, y
\end{lstlisting}

\begin{lstlisting}[style=pythonstyle, captionpos=b, caption={Function \texttt{Y} that applies the Pauli-Y operator}]
def Y(x: int, i: int, N: int):
    """
    Applies the Pauli-Y operator to the i-th site of the given state.
    
    Arguments:
        x (int): The input state as integer.
        i (int): The site index where the Pauli-Y operator is applied (0-based index).
        N (int): System size
        
    Returns:
        tuple: The coefficient and the resultant state.
    """

    # Convert integer to binary array
    state = bit(x, N)
        
    # Apply the Pauli-Y operator by flipping the i-th bit and multiplying by +/-i
    state[i] = 1 - state[i]
    
    # Determine the coefficient
    coef = (1 - 2 * state[i]) * 1.0j
        
    # Convert state to integer
    y = num(state)
            
    return coef, y
\end{lstlisting}

\begin{lstlisting}[style=pythonstyle, captionpos=b, caption={Function \texttt{Z} that applies the Pauli-Z operator}]
def Z(x: int, i: int, N: int):
    """
    Applies the Pauli-Z operator to the i-th site of the given state.
    
    Arguments:
        x (int): The input state as integer.
        i (int): The site index where the Pauli-Z operator is applied (0-based index).
        N (int): System size
        
    Returns:
        tuple: The coefficient and the resultant state.
    """
    
    # convert integer to binary array
    state = bit(x, N)
    
    # Apply the Pauli-Z operator: flip the sign if the i-th bit is 1    
    coef = 1 - 2 * state[i]
        
    return coef, x
\end{lstlisting}

\begin{lstlisting}[style=pythonstyle, captionpos=b, caption={Function \texttt{XX} that applies the Pauli-X operator to two sites}]
def XX(x: int, i: int, j: int, N: int):
    """
    Applies the Pauli-X operator to the i-th and j-th site of the given state
    
    Arguments:
        x (int): The input state as integer.
        i (int): The site index where the Pauli-X operator is applied (0-based index).
        j (int): The second site index for the Pauli-X operator.
        N (int): System size
        
    Returns:
        tuple: The coefficient and the resultant state.
    """
    
    # Convert integer to binary array
    state = bit(x, N)
        
    # Apply the Pauli-X operator by flipping the i-th and j-th bits
    state[i] = 1 - state[i]
    state[j] = 1 - state[j]
      
    coef = 1.0
    
    # Convert state to integer
    y = num(state) 
    
    return coef, y
\end{lstlisting}

\section{Codes for Chapter 6}

The following function corresponds to Algorithm \ref{alg:XXZ_states_magn},

\begin{lstlisting}[style=pythonstyle, captionpos=b, caption={Function \texttt{states generation} selects states with \texttt{n0} spins}]

def states_generation(N: int, n0: int) -> np.ndarray:
    """
    Selects states with n0 spins from the Hilbert space
    
    Arguments:
        N (int): system size
        n0 (int): number of up spins
        
    Returns:
        np.ndarray: list of states with n0 spins
    """
    
    if n0 > N:
        raise ValueError("n0 must be less than or equal to N")
        sys.exit()
        
    basis = []

    for i in range(2**N):
        if np.sum(bit(i, N)) == n0:
            basis.append(i)
            
    return np.array(basis)
    
\end{lstlisting}

\hspace{1cm}

The following function corresponds to Algorithm \ref{alg:find_position},

\begin{lstlisting}[style=pythonstyle, captionpos=b, caption={Function \texttt{find position} for binary search}]
def find_position(sorted_list: np.ndarray, target: int):
    """
    Perform bisection (binary) search on an ascending ordered list.

    Arguments:
        sorted_list (list): A list of elements in ascending order.
        target (any): The element to search for.

    Returns:
        int: The index of the target element if found, otherwise -1.
    """
    
    low = 0
    high = len(sorted_list) - 1

    while low <= high:
        mid = (low + high) // 2
        mid_val = sorted_list[mid]

        if mid_val == target:
            return mid
        elif mid_val < target:
            low = mid + 1
        else:
            high = mid - 1

    return -1
\end{lstlisting}

\hspace{1cm}

Finally, the following function corresponds to Algorithm \ref{alg:XXZ_Ham_magn},

\begin{lstlisting}[style=pythonstyle, captionpos=b, caption={Function \texttt{XXZ Ham U1} that generates the XXZ Hamiltonian matrix in a magnetization block}]
def XXZ_Ham_U1(N: int, J: float, delta: float, Sz: int) -> np.ndarray:
    """
    Generates the XXZ Hamiltonian matrix in a given magnetization block.
    
    Arguments:
        N (int): System size.
        J (float): Interaction strength.
        delta (float): Anisotropy parameter.
        Sz (int): Total magnetization.
        
    Returns:
        np.ndarray: The XXZ Hamiltonian matrix.
    """
    
    # Step 1: Compute n0 (assume N is even for simplicity)
    n0 = Sz + N // 2
        
    # Step 2: Generate all states in the magnetization block
    states = states_generation(N, n0)    
    D = len(states)
        
    # Step 3: Initialize the Hamiltonian matrix as a D x D zero matrix
    Ham = np.zeros((D, D))
    
    for x in range(D):  
        w = states[x]      
        b = bit(w, N)     
        
        for i in range(N):  
            j = np.mod(i + 1, N)  # j is the right nearest neighbor of i (PBC)
            
            if b[i] == b[j]:
                Ham[x, x] += -J * delta  # same configuration at i and j gives a negative contribution
            else:
                Ham[x, x] += J * delta  
                b_new = flip(b, i, j)    
                y = num(b_new)           
                z = find_position(states, y)  
                Ham[x, z] += -J         
                
    return Ham
\end{lstlisting}

\chapter{Classical and Quantum Phase Transitions}\label{ch:cpt}

\section{Phenomenology of classical phase transitions}

\begin{figure}\label{fig:magnetic_domain}
    \centering
\includegraphics[width=0.75\linewidth]{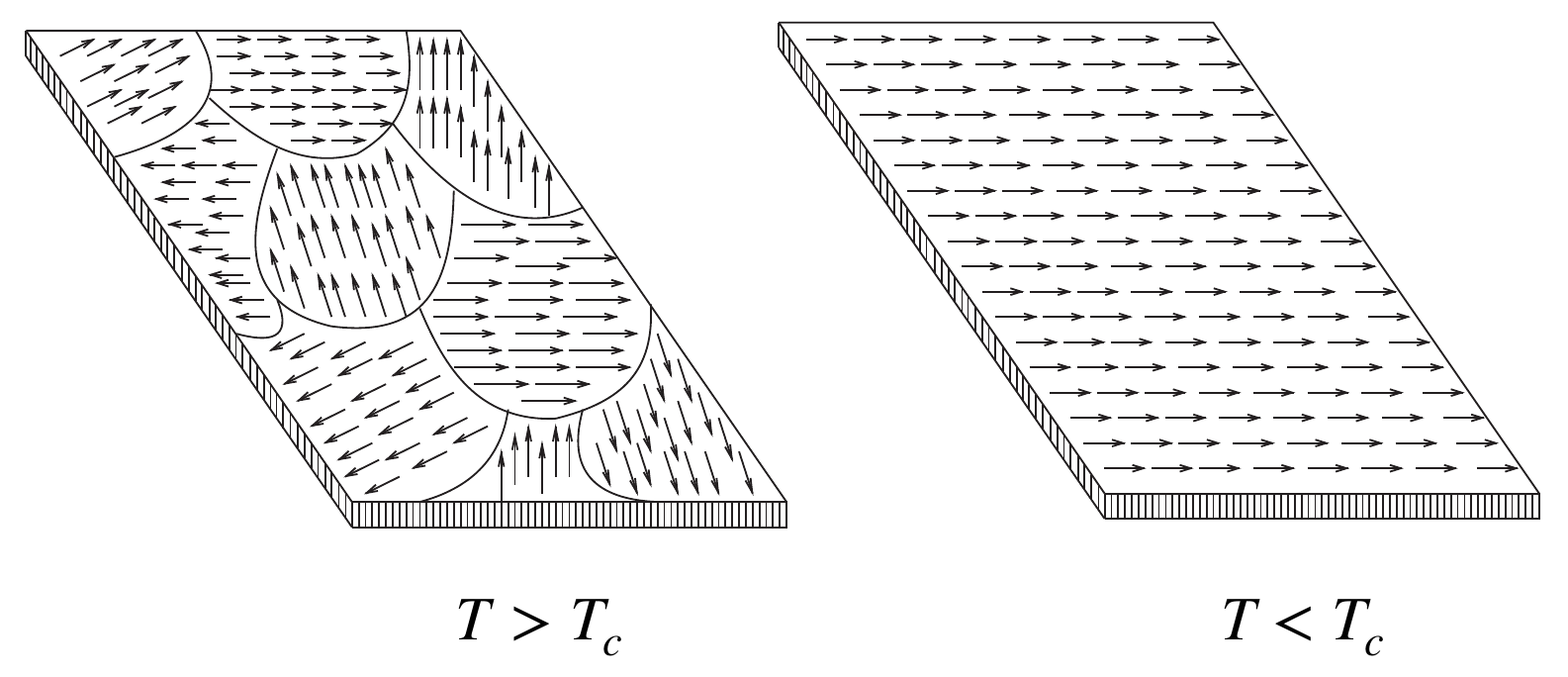}
\caption{Ordered and disordered phases of a ferromagnetic system.}
\end{figure}

Atoms in certain materials possess an intrinsic magnetic dipole moment, arising from the spin and orbital motion of their electrons. In most materials, however, these magnetic dipoles are randomly oriented, causing them to cancel each other out and resulting in no net magnetic field.

However, in some materials, such as iron or cobalt, interactions between atomic dipoles create a noticeable overall magnetic field. This occurs in \textit{ferromagnetic materials} when the temperature is below a specific critical value called the Curie temperature ($T_{c}$), which varies by material. At Curie temperature, these materials undergo a \textit{phase transition}, changing their physical properties. In this case, the change is the spontaneous alignment of microscopic dipoles, resulting in a macroscopic magnetic field. This phase transition occurs due to two competing factors. 

\begin{itemize}
\item \textbf{Energy Minimization} --
In ferromagnetic materials, the magnetic dipoles (or spins) of each atom tend to align to minimize the system's total energy. This alignment occurs due to quantum effects, but it can also be modelled classicaly. If energy minimization were the sole governing principle, all these dipoles (spins) would align perfectly, causing many materials to exhibit a strong magnetic field. However, this is not observed in practice because a competing mechanism—thermal fluctuations—comes into play, disrupting perfect alignment.

\item \textbf{Entropy Maximisation} --
Among the many possible configurations of a system, those where spins align in the same direction are rare. Unless it requires a lot of energy to reorient neighboring spins, there are far more configurations with randomly oriented spins than with fully aligned spins. Entropy $S $, which measures disorder, is given by Boltzmann's formula:
\begin{equation}\label{eq:entropy}
    S(E) = k \log \omega(E)
\end{equation}
where $\omega(E)$is the number of states at energy $E $, and $k$is the Boltzmann constant. If systems only sought to maximize disorder, spontaneous magnetization would never occur.

\end{itemize}
As illustrated by the example of magnetic dipoles, macroscopic physical systems with a vast number of degrees of freedom are governed by two competing tendencies: one driving the system toward order to minimize energy and another favoring disorder to maximize entropy. For a genuine competition between these tendencies to emerge, an additional critical factor must be considered---the system's temperature. The influence of temperature $T$ in this context is dictated by the principles of statistical mechanics.

\subsection{Classification of the phase transitions}
In modern physics, phase transitions are classified into two main types: first-order and second-order phase transitions \footnote{There are also ``infinite-order phase transitions'', which are continuous but do not break any symmetries. A well-known example is the Kosterlitz–Thouless transition in the two-dimensional XY model.}.

\begin{itemize}
    \item \textbf{First-order phase transitions} involve a \textit{latent heat}. This means that at the transition point, the system absorbs or releases a fixed amount of energy (heat), while the temperature remains constant. These transitions are marked by a finite correlation length, leading to a mixed-phase state where parts of the system have completed the transition, while others have not. For example, when water cools to its freezing point $T_f$, it does not instantly turn to ice but instead forms a mixture of water and ice. The latent heat indicates a significant structural change in the material: above $T_f$, water molecules move freely in a disordered manner, while below $T_f$, they are arranged in an orderly crystal lattice of ice. Other examples of first-order transitions include Bose–Einstein condensation.
    \item \textbf{Second-order phase transitions}, also known as continuous phase transitions, do not involve latent heat. Instead, they are characterized by a \textit{divergence of the correlation length at the critical point}. Examples include the ferromagnetic transition, superconductors, and the superfluid transition. 
\end{itemize}

\section{Brief review of Statistical Physics}
One of the most significant advancements in nineteenth-century physics was the discovery of the understanding of the statistical laws governing many-body systems at thermal equilibrium. These laws form the foundation of statistical mechanics and have become a cornerstone of modern theoretical physics.

Let $\pmb{\sigma}$ represent a generic state (configuration) of a physical system, such as the orientation of each magnetic dipole. For a system at thermal equilibrium with a large number $N$ of spins / particles, the probability $p(\pmb{\sigma})$ of observing the system in a given configuration $\pmb{\sigma}$ is given by the Boltzmann law
\begin{equation}
    p(\pmb{\sigma}) = \frac{e^{-\beta H(\pmb{\sigma})}}{Z} \, ,
\end{equation}
where $H(\pmb{\sigma})$ is the energy of the configuration $\pmb{\sigma}$, $\beta = \frac{1}{kT}$ with $k$ being the Boltzmann constant, and $T$ the absolute temperature. The partition function $Z$ is defined as
\begin{equation}
Z = \sum_{\pmb{\sigma}} e^{-\beta H(\pmb{\sigma})} \, ,
\end{equation}
and normalizes the Boltzmann probabilities, also encoding all relevant physical information about the system at equilibrium. If we rewrite this sum by grouping together all configurations $\pmb{\sigma}$ that share the same energy $E = H(\pmb{\sigma})$, we obtain:
\begin{equation}
Z = \sum_{E} \omega(E) e^{-\beta E} \, ,
\end{equation}
where $\omega(E)$ is the total number of configuration with energy $E$. Now, by using the definition of entropy (Eq.\ref{eq:entropy}), we get
\begin{equation}
Z = \sum_{E}  e^{-\beta E + \log \omega(E)}
= \sum_{E}  e^{-\beta \left(E - T S(E) \right)} \, .
\end{equation}
The \textit{free energy} $F$ is related to $Z$ by:
\begin{equation}
Z \equiv e^{-\beta F(N, \beta)} \, ,
\end{equation}
where $F$ is given by:
\begin{equation}
F = U - TS \, .
\end{equation}
Here, $U = \langle E \rangle$ is the internal energy, and $S$ is the entropy. The thermodynamic relations for $U$ and $S$ are:
\begin{equation}
\langle U \rangle = \frac{\partial (\beta F)}{\partial \beta}
, \quad
\langle S \rangle = \beta^2 \frac{\partial F}{\partial \beta} \, .
\end{equation}
These equations highlight how the balance between energy and entropy changes with temperature, potentially leading to phase transitions at a critical temperature $T_c$.

\section{The Peierls argument}

The Peierls argument provides a way to understand phase transitions by considering the stability of ordered states against the formation of domain walls. Domain walls are boundaries separating regions with different spin orientations. If creating these walls costs too much energy, the system prefers to remain ordered, leading to a phase transition at a certain temperature.

\subsection{The Ising Model in 1D}

In the one-dimensional classical Ising model, spins $\spin_i = \pm 1$ are arranged on a linear 1D lattice. The Hamiltonian of the system is given by: $H = -J \sum_{i} \spin_i \spin_{i+1}$, where $J > 0$ is the coupling constant.

\begin{equation}\label{eq:brickwall}
\includegraphics[width=0.45\linewidth, valign=c]{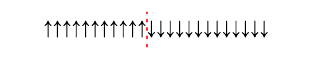}
\end{equation}

\begin{enumerate}
    \item \underline{Domain Wall Energy}: 
    \begin{itemize}
        \item In 1D, a domain wall is a boundary between two spins pointing in opposite directions.
        \item The energy cost to create one domain wall is $2J$ because it breaks two bonds.
    \end{itemize}
    
    \item \underline{Entropy Consideration}: 
    \begin{itemize}
        \item The number of possible positions for a domain wall in a system with $N$ spins is $N$.
        \item The entropy associated with placing a domain wall is $k_B \ln(N) $.
    \end{itemize}
    
    \item \underline{Free Energy}: 
    \begin{equation}
    \Delta F = 2J - k_B T \ln(N).
    \end{equation}
\end{enumerate}

At any finite temperature $T$, the entropy term $k_B T \ln(N)$ will eventually dominate for large $N$, making $\Delta F$ negative. This implies that creating domain walls becomes favorable, leading to a disordered phase. Therefore, no long-range order (ferromagnetic order) is expected at any finite temperature in the 1D Ising model.

\subsection{The Ising Model in 2D}

In the 2D Ising model, spins are arranged on a lattice, and the Hamiltonian is $H = -J \sum_{\langle i,j \rangle} \spin_i \spin_j$, 
where the sum runs over nearest neighbor spins.

\begin{equation}\label{eq:brickwall}
\includegraphics[width=0.45\linewidth, valign=c]{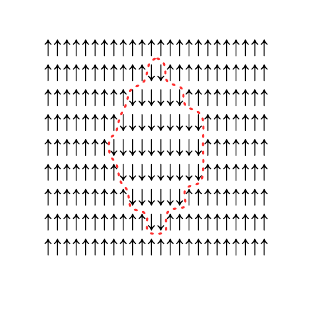}
\end{equation}

\begin{enumerate}
    \item \underline{Domain Wall Energy}: 
    \begin{itemize}
        \item In 2D, a domain wall is a line of flipped spins separating regions with different orientations.
        \item The energy cost of a domain wall is proportional to its length $L $, so $\Delta E \approx 2JL $.
    \end{itemize}
    
    \item \underline{Entropy Consideration}: 
    \begin{itemize}
        \item The number of ways to arrange a domain wall of length $L$ is exponential in $L$, $\exp(L)$.
        \item The entropy associated with a domain wall of length $L$ is $k_B L \ln 2 $.
    \end{itemize}
    
    \item \underline{Free Energy}: 
    \begin{equation}
    \Delta F = 2JL - k_B T L \ln 2.
    \end{equation}
\end{enumerate}

At low temperatures, the energy term $2JL$ dominates, making $\Delta F$ positive, and domain walls are not favorable. This leads to long-range order and a finite temperature transition in the 2D Ising model. However, as the temperature increases, the entropy term $k_B T L \ln 2$ starts to dominate, and $\Delta F$ can become negative, favoring domain wall creation and leading to a disordered phase. This transition occurs at a critical temperature $T_c$.

We can estimate the critical temperature $T_c$ for the 2D Ising model by equating the energy and entropy contributions:
\begin{equation}
2JL \approx k_B T_c L \ln 2 
\end{equation}
which gives $T_c \approx \frac{2J}{k_B \ln 2}$. While this is a rough estimation, it captures the idea that a phase transition occurs when the thermal energy is comparable to the energy cost of creating domain walls.

\section{Exact solution of the Ising model in 1D}
The partition function of the 1D Ising model can be computed exactly using the transfer matrix approach. This method is a powerful tool to solve statistical mechanics problems, particularly in one dimension.

We consider a 1D Ising Hamiltonian 
\begin{equation}
H(\pmb{\sigma}) = -J \sum_{i=1}^N \spin_i \spin_{i+1} - h \sum_{i=1}^{N} \spin_i \, ,
\end{equation}
where $J$ is the coupling constant and $h$ is the external magnetic field. For simplicity, we assume periodic boundary conditions: $\spin_{N+1} = \spin_1$. The partition function $Z$ is given by:
\begin{equation}
Z = \sum_{\pmb{\sigma}} \exp\left(-\beta H\right)
\end{equation}
where $\beta = \frac{1}{k_B T}$, $k_B$ is the Boltzmann constant, and $T$ is the temperature. Now, we rewrite the Hamiltonian in terms of a certain $2 \times 2$ matrix $\mathcal{T}$ named \textit{transfer matrix}. The transfer matrix elements are defined as:
\begin{equation}
\mathcal{T}(\sigma_i, \sigma_{i+1}) = \exp\left[ \beta (J \sigma_i \sigma_{i+1} + \frac{h}{2}(\sigma_i + \sigma_{i+1})) \right]
\end{equation}
thus the matrix reads
\begin{equation}
\mathcal{T} = \begin{pmatrix}
\exp(\beta J + \beta h) & \exp(-\beta J) \\
\exp(-\beta J) & \exp(\beta J - \beta h)
\end{pmatrix}
\end{equation}

The partition function $Z$ is then given by:
\[
Z = \text{Tr}(\mathcal{T}^N)
\]

To compute $\mathcal{T}^N$, we first diagonalize $\mathcal{T}$. Let $\lambda_1$ and $\lambda_2$ be the eigenvalues of $\mathcal{T}$. The eigenvalues are found by solving the characteristic equation:
\[
\det(\mathcal{T} - \lambda I) = 0
\]

The eigenvalues $\lambda_1$ and $\lambda_2$ are:
\begin{equation}
\lambda_{1,2} = \exp(\beta J) \cosh(\beta h) \pm \sqrt{\exp(2 \beta J) \sinh^2(\beta h) + \exp(-2 \beta J)} \, .
\end{equation}
Then $\mathcal{T}^N$ can be written as:
\[
\mathcal{T}^N = M \begin{pmatrix}
\lambda_1^N & 0 \\
0 & \lambda_2^N
\end{pmatrix} M^{-1}
\]
where $P$ is the matrix of eigenvectors.

Thus, the partition function is:
\begin{equation}
Z = \lambda_1^N + \lambda_2^N \,.
\end{equation}
For large $N$, the largest eigenvalue $\lambda_1$ dominates:
\begin{equation}
Z \approx \lambda_1^N \, .
\end{equation}
In one dimension, the largest eigenvalue $\lambda_1$ always dominates the partition function for any finite temperature. As a result, the free energy 
\begin{equation}
    F = -k_B T \ln Z \approx - N k_B T \log \lambda_1
\end{equation}
is an analytical function of the temperature $T$, for all $T>0$. This analyticity implies that there is no phase transition at any finite temperature. Thus, the analytical findings using the transfer matrix approach corroborate what is predicted by the Peierls argument regarding the absence of a phase transition in 1D.

\section{Phenomenology of quantum phase transitions}

A classical phase transition is characterized by a dramatic change in the properties of a system, driven by thermal fluctuations. In contrast, a quantum phase transition (QPT) occurs at zero temperature and is driven by \textit{quantum fluctuations}~\cite{Sachdev_2011}. These fluctuations are influenced by a physical parameter $h$, such as an internal interaction or an external field, which is part of the system's Hamiltonian $\hat{H}(h)$. 

\begin{figure}[ht!]
    \centering
\includegraphics[width=0.65\linewidth]{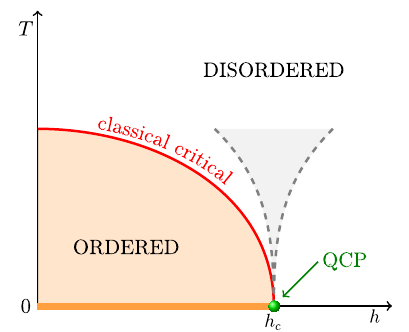}
\caption{\label{fig:phase_transition}
Sketch of a quantum phase diagram where transitions are driven by temperature $T$ and a physical parameter $h$. 
}
\end{figure}

While classical phase transitions are characterized by singularities in the system's thermodynamic functions, quantum phase transitions are distinguished by \textit{singular (nonanalytic) features in the many-body ground state} $|\Psi_0(h) \rangle$~\cite{Sachdev_2011}. This leads to profound differences in the system's macroscopic properties on both sides of the transition.

At finite but low temperatures, quantum fluctuations compete with thermal fluctuations. This competition causes the quantum transition to influence a small region of the $(T, h)$ phase diagram, where $T$ is the temperature and $h$ is the Hamiltonian parameter driving the quantum transition. As the temperature $T$ increases, the system transitions from quantum critical behavior to classical behavior (provided that a classical phase transition exists). Figure~\ref{fig:phase_transition} illustrates a typical phase diagram in the $(T, h)$ plane. At high temperatures, the system is disordered. Near the classical phase transition (indicated by the red line), thermal fluctuations dominate. As the temperature decreases, this transition line approaches the QCP (marked by the green circle). The light gray region represents the crossover where quantum and thermal fluctuations are comparable. Finally, the orange line at $T=0$ shows the ordered pure quantum phase.

\subsection{Level crossing}

\begin{figure}
    \centering
\includegraphics[width=\linewidth]{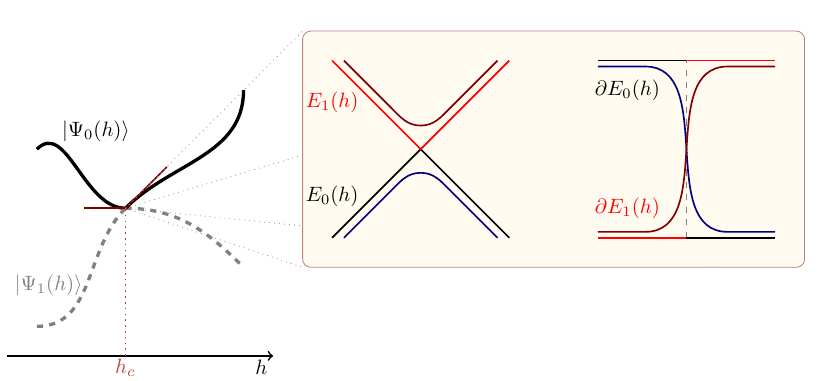}
\caption{\label{fig:level_crossing}
Pictorial representation of the ground-state path in the Hilbert space. In such example, the point where the curve is no more derivable represent a QCP. In the inset we inspect some properties of the system near the QCP: in the left figure, the blue and the dark red curves are respectively the energies of the ground state and the first-excitation state for a gapped system. Dark and light red curves represent the energy behavior for a gapless system; notice how level crossing in the energy (gapless system) corresponds to a discontinuity on the first derivative (right part of the inset); nevertheless in a more general setup, e.g. for a finite system, where level crossing is forbidden, the discontinuity is rounded off.
}
\end{figure}

Let us delve deeper into the details of a QPT. If the system's Hamiltonian is $\hat{H}(h)$, the ground state $|\Psi_0(h) \rangle$ will vary with $h$. The nonanalytic behavior at the quantum critical point (QCP) can manifest through the expectation value 
\begin{equation}
  \langle \hat{O} \rangle(h) = \langle \Psi_0(h) | \hat{O} | \Psi_0(h)\rangle  \, ,
\end{equation}
of a given observable $\hat{O}$. For example, in the paramagnetic to ferromagnetic transition of the quantum Ising model, the total longitudinal magnetization $\hat{M}_z = \sum_{i} \pauliz_i$ exhibits nonanalytic behavior at the transition point. Interestingly, if $\hat{O}$ is the Hamiltonian operator itself, its expectation value will simply be the energy of the ground state $E_0(h)$.

When the system is far from the QCP, the ground state $|\Psi_0(h) \rangle$ traces a smooth path in the Hilbert space as a function of $h$. However, as the QCP $h_c$ is approached (as shown in Figure~\ref{fig:level_crossing}), this path becomes nonanalytic because the ground state may overlap with excited states. 

As the ground-state energy $E_0(h)$ approaches the energy of an excited state $E_1(h)$, the system undergoes a fundamental change in its properties. Specifically, in the vicinity of $h_c$, the level crossing will manifest as a discontinuity in the first derivative of $E_0(h)$ (see the inset of Figure~\ref{fig:level_crossing}). 

\begin{figure}
    \centering
\includegraphics[width=0.75\linewidth]{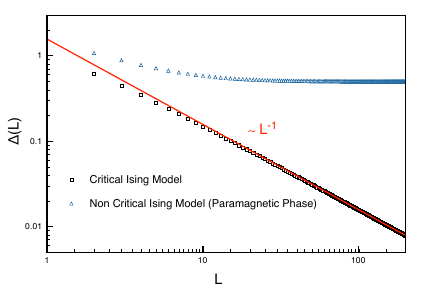}
\caption{\label{fig:ising_gap}
Finite-size scaling of the gap $\Delta$ for the Ising model. The system is gapped in the paramagnetic phase. Otherwise, at the critical point, the gap closes as $L^{-1}$.
}
\end{figure}

In finite-dimensional systems, level crossings are typically forbidden unless the system is in the thermodynamic limit. As illustrated in Figure~\ref{fig:ising_gap} for the Ising model, in the thermodynamic limit, the energy gap remains finite away from the critical point but closes as $N^{-1} $ exactly at $h_c $, where $N$ is the system size.

For second-order QPTs, the energy gap $ \Delta(h) = |E_1(h) - E_0(h)| $ serves as the key energy scale near the transition. Near $ h_c $, the gap follows the scaling relation:
\begin{equation}
    \Delta(h) \sim J |h - h_c|^{z \nu}
\end{equation}
where $J$ is a microscopic energy scale, $z$ is the so-called \textit{dynamical critical exponent}, and $\nu$ is the correlation length exponent. The correlation length $ \xi$ diverges at $h_c$ as:
\begin{equation}
\xi \sim a |h - h_c|^{-\nu}, \quad \text{or} \quad \xi \sim \Delta^{-1/z}
\end{equation}
with $a$ being a characteristic length (e.g.\ lattice spacing). The characteristic time scale of a quantum system, given by the inverse of the minimum gap $ \tau \sim \Delta^{-1} $, parallels the correlation time in classical systems.

The adiabatic theorem states that a quantum system can be driven adiabatically if the Hamiltonian changes slowly compared to the system's relaxation time. Near the critical point, as the gap closes, this adiabatic condition breaks down. Thus, close to $ h_c $, the system may not follow an adiabatic path, making the choice of an appropriate temporal protocol crucial for guiding the system through a QPT.

\subsubsection{A simple example: Landau-Zener}

Consider the Hamiltonian $\hat{H}(\lambda)$ describing a single spin-$1/2$ in the presence of a longitudinal magnetic field (along the $z$-axis) proportional to $\lambda$, combined with a fixed transverse field (along the $x$-axis) of strength $\Delta$:
\begin{equation}
\hat{H}(\lambda) = \Delta \paulix + \lambda \pauliz \, .
\end{equation}
When $\Delta=0$, $\hat{H}$ describes a simple two-level system with energies $\pm \lambda$. The term $\Delta$ instead mixes the two levels, leading to the spectrum
\begin{equation}
\hat{H}(\lambda) | \pm \rangle = E_{\pm} | \pm \rangle, \quad E_{\pm} = \pm \sqrt{\lambda^2 + \Delta^2} \,
\end{equation}
and corresponding eigenstates
\begin{equation}
| + \rangle = 
\begin{pmatrix}
\cos(\theta/2) \\
\sin(\theta/2)
\end{pmatrix}, \quad
| - \rangle = 
\begin{pmatrix}
-\sin(\theta/2) \\
\cos(\theta/2)
\end{pmatrix},
\end{equation}

where

\begin{equation}
\cos \theta = \frac{\lambda}{\sqrt{\lambda^2 + \Delta^2}}, \quad \sin \theta = \frac{\Delta}{\sqrt{\lambda^2 + \Delta^2}}.
\end{equation}

\begin{figure}[h]
    \centering
    \includegraphics[width=0.5\textwidth]{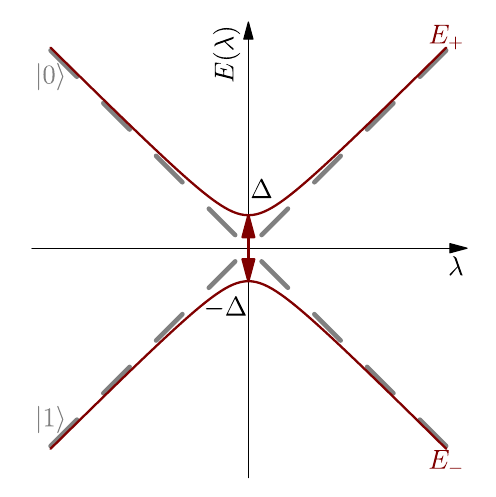}
    \caption{Spectrum of the spin Hamiltonian. In the absence of the transverse field, the two levels $|0\rangle$ and $|1\rangle$ cross at $\lambda = 0$. For any other value of $\Delta$, the level crossing is avoided. The two levels $E_{\pm}$, corresponding to the eigenstates $| \pm \rangle$, are separated by a gap $2\Delta$.}
    \label{fig:spectrum}
\end{figure}

We observe that the topology of the spectrum changes between the cases $\Delta = 0$ and $\Delta \neq 0$; in the former case, there is a level crossing which disappears for any finite transversal field. There is a loss of analyticity in the Hamiltonian's spectrum for a particular value of the parameters. Note, however, that when $\Delta \neq 0$, transition between the positive and negative energy levels is impossible.

Now assume that $\lambda = \lambda(t)$ becomes a time-dependent parameter:

\begin{equation}
\lambda(t) = vt,
\end{equation}

where $v$ characterizes the variation of $\lambda$ over time (we take $\hbar = 1$). The time dependence of the Hamiltonian allows the emergence of a new effect, the transition between the two levels $E_{+}$ and $E_{-}$, absent in the static case. This is the effect discovered by Landau, Zener, Majorana, and Stückelberg (1932), independently. They found that the transition probability between the two adiabatic levels $E_{\pm}$ (for the linear time dependency of $\lambda$) is exponentially small:

\begin{equation}
p_{+-} = e^{-\pi \Delta^2 / v}.
\end{equation}

This expression gives the transition probability at infinite time ($t \rightarrow \infty$), when the initial state was on one of the $E_{\pm}$ levels, at $t \rightarrow -\infty$.

\section{Transverse field Ising Model}
To illustrate the quantum phase transition, we consider the transverse field Ising model, whose Hamiltonian reads (see Section \ref{sec:ising_quantum})
\begin{equation}
    \hat{H} = - \sum_{i=1}^{N-1} \pauliz_{i} \pauliz_{i+1} - h \sum_{i=1}^N  \paulix_{i} \, ,
\end{equation}
Notice that for simplicty we set the ferromagnetic interaction $J$ to $1$ (and we omit the longitudinal field). When the magnetic field $h$ is much stronger than the interaction strength $J$, $h \gg 1$, the system’s spins align with the field in the $x$ direction. Consequently, the ground state $|\Psi_{0} \rangle$ becomes a simple tensor product of eigenstates of the $\paulix$ operator (see Eq.\ref{eq:eigenstates_x}):
\begin{equation}
|\Psi_{0} \rangle \stackrel{h \gg 1}{\longrightarrow} \bigotimes_{i} \left( \frac{\ket{\uparrow} + \ket{\downarrow}}{\sqrt{2}} \right)_i \, .
\end{equation}
In this state, the spins are effectively uncorrelated, and the correlation function in the $z$ direction is simply:
\begin{equation}
\langle \Psi_{0} | \pauliz_{i} \pauliz_{j} | \Psi_{0} \rangle \bigg|_{h \to \infty} = \delta_{ij} \, .
\end{equation}
Conversely, when the ferromagnetic interaction strength dominates, i.e.\ for $h=0$, spins tend to align in the $\hat{z}$ direction, leading to two degenerate ferromagnetic ground states
\begin{equation}
\bigotimes_{i} \ket{\uparrow}_i \quad \text{and} \quad \bigotimes_{i} \ket{\downarrow}_i \, .
\end{equation}
When the system is in one of these two states, it exhibits long range (yet trivial) correlations. In fact the correlation function in the $z$ direction reads
\begin{equation}
   \langle \Psi_{0} | \pauliz_{i} \pauliz_{j} | \Psi_{0} \rangle \bigg|_{h = 0} = 1 \, .
\end{equation}
A QPT  occurs at an intermediate value of the field $h \sim J = 1$, where the system transitions between these two regimes. The critical field is denoted as $h_c$, and for the transverse field Ising model takes the value $h_c=J=1$. The transition can be observed by studying the behavior of the $z$-correlation function 
\begin{equation}
\lim_{|i-j| \to \infty} \langle \Psi_{0} | \pauliz_{i} \pauliz_{j} | \Psi_{0} \rangle \, ,
\end{equation}
which indicates long-range order in the ferromagnetic phase ($h < h_c$) and short-range correlations in the paramagnetic phase ($h > h_c$). Another way of detecting the transition is to study the value of the order parameter of the model, namely the longitudinal magnetization 
\begin{equation}
    \langle \Psi_{0} | \pauliz_{i} | \Psi_{0} \rangle \, .
\end{equation}

\bibliographystyle{unsrt}
\bibliography{bib}

\begin{thebibliography}{10}

\bibitem{Sandvik_2010}
Anders~W. Sandvik.
\newblock {Computational Studies of Quantum Spin Systems}, 2010.
\newblock \url{https://arxiv.org/abs/1101.3281}.

\bibitem{Troyer_2015}
Matthias Troyer.
\newblock {Computational Quantum Physics}.
\newblock \url{https://share.phys.ethz.ch/~alps/cqp.pdf}, 2015.
\newblock Accessed: 2024-07-01.

\bibitem{paffuti2009quantum}
K.~Konishi and G.~Paffuti.
\newblock {\em {Quantum Mechanics: A New Introduction}}.
\newblock OUP Oxford, 2009.

\bibitem{sakurai2011modern}
J.J. Sakurai and J.~Napolitano.
\newblock {\em {Modern Quantum Mechanics}}.
\newblock Addison-Wesley, 2011.

\bibitem{Huang1963StatisticalM2}
Kerson Huang.
\newblock {\em {Statistical Mechanics, 2nd Edition}}.
\newblock John Wiley and Sons, 1963.

\bibitem{Tong_2012}
David Tong.
\newblock {Lectures on Statistical Physics}.
\newblock \url{https://www.damtp.cam.ac.uk/user/tong/statphys.html}, 2012.
\newblock Accessed: 2024-07-01.

\bibitem{Sachdev_2011}
Subir Sachdev.
\newblock {\em {Quantum Phase Transitions}}.
\newblock Cambridge University Press, 2 edition, 2011.

\bibitem{Nielsen_chuang_2010}
Michael~A. Nielsen and Isaac~L. Chuang.
\newblock {\em {Quantum Computation and Quantum Information: 10th Anniversary Edition}}.
\newblock Cambridge University Press, 2010.

\bibitem{Santoro_2024}
Giuseppe E.Santoro.
\newblock {Introduction to Quantum Computation and Information}.
\newblock \url{https://cm.sissa.it/phd/courses/seminar-series-quantum-computation-and-information}, 2024.
\newblock Accessed: 2024-07-01.

\bibitem{SCHOLLWOCK201196}
Ulrich Schollwöck.
\newblock {The density-matrix renormalization group in the age of matrix product states}.
\newblock {\em Annals of Physics}, 326(1):96--192, 2011.
\newblock January 2011 Special Issue.

\bibitem{Collura2024}
Mario Collura, Guglielmo Lami, Nishan Ranabhat, and Alessandro Santini.
\newblock {\em Tensor Network Techniques for Quantum Computation}.
\newblock SISSA Medialab S.r.l., 2024.

\bibitem{enwiki:1221833336}
{Wikipedia contributors}.
\newblock Quadratic unconstrained binary optimization --- {Wikipedia}{,} the free encyclopedia.
\newblock \url{https://en.wikipedia.org/w/index.php?title=Quadratic_unconstrained_binary_optimization&oldid=1221833336}, 2024.
\newblock [Online; accessed 6-July-2024].

\bibitem{enwiki:1220838427}
{Wikipedia contributors}.
\newblock Characteristic polynomial --- {Wikipedia}{,} the free encyclopedia.
\newblock \url{https://en.wikipedia.org/w/index.php?title=Characteristic_polynomial&oldid=1220838427}, 2024.
\newblock [Online; accessed 14-June-2024].

\bibitem{Wei_2022}
David Wei, Antonio Rubio-Abadal, Bingtian Ye, Francisco Machado, Jack Kemp, Kritsana Srakaew, Simon Hollerith, Jun Rui, Sarang Gopalakrishnan, Norman~Y. Yao, Immanuel Bloch, and Johannes Zeiher.
\newblock {Quantum gas microscopy of Kardar-Parisi-Zhang superdiffusion}.
\newblock {\em Science}, 376(6594):716–720, May 2022.

\bibitem{enwiki:1228829178}
{Wikipedia contributors}.
\newblock Binary search --- {Wikipedia}{,} the free encyclopedia.
\newblock \url{https://en.wikipedia.org/w/index.php?title=Binary_search&oldid=1228829178}, 2024.
\newblock [Online; accessed 13-July-2024].

\bibitem{Hastings_2007}
M~B Hastings.
\newblock {An area law for one-dimensional quantum systems}.
\newblock {\em Journal of Statistical Mechanics: Theory and Experiment}, 2007(08):P08024–P08024, August 2007.

\end{thebibliography}

\end{document}